\documentclass{article}
\pdfoutput=1
\usepackage{lscape}
\usepackage{enumerate}
\usepackage{amsfonts}
\usepackage{amsmath}
\usepackage{textcomp}
\usepackage{amssymb}
\usepackage{color}
\usepackage{graphicx}
\usepackage{pdflscape}
\usepackage{afterpage}
\usepackage[matrix,arrow,curve]{xy}
\usepackage{array}
\usepackage{indentfirst}

\def\be{\begin{equation}}
\def\ee{\end{equation}}

\makeatletter
\renewcommand*{\@cite@ofmt}{\bfseries\hbox}
\makeatother

\hoffset= - 1.2in         

\begin{document}

\title{\vspace{0.1cm}{\Large {\bf  Quantum Racah matrices up to level 3 and multicolored link invariants}\vspace{.2cm}}
\author{{\bf C. Bai$^a$}, \ {\bf J. Jiang$^b$}, \ {\bf J. Liang$^a$}, \ {\bf A. Mironov$^{c,d,e}$}, \\
{\bf A. Morozov$^{d,e}$}, \ {\bf An. Morozov$^{d,e}$}, \ {\bf A. Sleptsov$^{d,e,f}$}
}
\date{ }
}

\maketitle

\vspace{-5.5cm}

\begin{center}
\hfill FIAN/TD-31/17\\
\hfill IITP/TH-02/18\\
\hfill ITEP/TH-02/18\\
\end{center}

\vspace{4.2cm}

\begin{center}

$^a$ {\small {\it Chern Institute of Mathematics and LPMC, Nankai University, Tianjin 300071, China}}\\
$^b$ {\small {\it School of Math \& Physics, Ningde Normal University, Ningde 352100, China}}\\
$^c$ {\small {\it Lebedev Physics Institute, Moscow 119991, Russia}}\\
$^d$ {\small {\it ITEP, Moscow 117218, Russia}}\\
$^e$ {\small {\it Institute for Information Transmission Problems, Moscow 127994, Russia}}\\
$^f$ {\small {\it Laboratory of Quantum Topology, Chelyabinsk State University, Chelyabinsk 454001, Russia }}
\end{center}

\vspace{1cm}

\begin{abstract}
This paper is a next step in the project of systematic description of colored knot and link invariants started in \cite{MMfing,Rama2}. In this paper, we managed to explicitly find
the \underline{inclusive} Racah matrices, i.e. the whole set of mixing matrices in channels $R_1\otimes R_2\otimes R_3\longrightarrow Q$ with all possible $Q$, for $|R|\leq 3$. The calculation is made possible by use of the highest weight method. The result allows one to evaluate and investigate colored polynomials for arbitrary 3-strand knots and links and to check the corresponding eigenvalue conjecture. Explicit answers for Racah matrices and colored polynomials for 3-strand knots up to 10 crossings are available at \cite{knotebook}. Using the obtained inclusive Racah matrices, we also calculated the \underline{exclusive} Racah matrices with the help of trick earlier suggested in the case of knots. This method is proved to be effective and gives the exclusive Racah matrices earlier obtained by another method.
\end{abstract}

\vspace{.5cm}

\tableofcontents

\section{Introduction}

Knot and link invariants are now widely studied objects. In part, this is due to their many connections to other areas of physics. One of the most important connections was first realized by E.Witten \cite{Wit} for a particular set of knot polynomials, Jones polynomials \cite{Jones}, which, as he claimed, were equal to the Wilson-loop averages of the Chern-Simons theory with the gauge group $SU(2)$ \cite{CS}. This was generalized to much wider classes of polynomials and consequently gauge groups, that is, to the HOMFLY-PT polynomials \cite{HOMFLY} and the gauge group $SU(N)$ or the Kauffman polynomials \cite{Kauf} and the gauge group $SO(N)$.

What makes knot/link invariants interesting from the point of view of quantum field theories is that these are exact answers for the averages, i.e. they are calculated in their exact form without using any kind of perturbation theory. This makes them a rare example of such observables in quantum field theories.

Also widely known are connections between knot polynomials and transformations of conformal blocks in 2D CFT \cite{Wit,inds}. Due to this relation, the exclusive Racah, which we discuss later in this paper can be used to describe modular transformations of the corresponding conformal blocks.
Quite important are connections with topological strings \cite{OV} which in particular implies certain integrality conjectures for knot and link invariants (see \cite{LMOV}), and the results of the present paper could help in further checking this conjecture for links and consequently check relations with topological strings.
Another important connection is possible topological quantum computing \cite{QC}. It can be based on anyons, which are quasi-particles interacting with the Chern-Simons action. The observables (programs) for this quantum computer should be interpreted as some knot invariants. Thus calculations made in the present paper can provide a further insight in how this quantum computer can work.

\bigskip

The present paper is a continuation of the long program devoted to studies of knot invariants \cite{MMfing,Rama2}. In our previous papers, we managed to calculate inclusive and exclusive Racah matrices needed for calculations of knot polynomials in representations up to size four. In the present paper, we further generalize these results to include not only knots but links with distinct representations on distinct components. First, we calculate the inclusive Racah matrices and then, using these matrices, we also calculate the exclusive Racah matrices with the help of the trick earlier suggested in the case of knots \cite{MMMSint}. This gives a check of this method, which is proved to be effective and allows us to reproduce the exclusive Racah matrices earlier obtained by another method \cite{RacahRama}.

The present paper is organized as follows. Section \ref{s.rac} is devoted to the description of Racah matrices with s.\ref{incrac} dealing with inclusive Racah matrices and s.\ref{exrac} with exclusive Racah matrices. Section \ref{incrac} also deals with the definition of $\mathcal{R}$-matrices. Section \ref{s.raccalc} describes the methods to calculate the Racah matrices with s.\ref{s.exin} providing details on how to find the exclusive matrices from the inclusive ones, while s.\ref{hwc} recalls the highest weight calculus method for inclusive Racah matrices. Section \ref{s.raclist} provides the list of all Racah matrices for representations up to size 3. Section \ref{s.signs} describes how the signs of $\mathcal{R}$-matrices eigenvalues are defined. Section \ref{s.eig} recalls the eigenvalue hypothesis for the Racah matrices which immediately provides answers for a number of Racah matrices. As an example of application of the obtained Racah matrices, we consider three essentially different links that admit 3-stand braid representation: the Whitehead link, the three-component Borromean rings, and link $L7a3$, which has two components, one of them being not unknot. Section \ref{s.HOMFLY} introduces the links invariants that we evaluate, with the whole list of answers contained in Appendix \ref{a.pol}. Section \ref{s.exex} provides an example of evaluating the exclusive Racah matrix from the inclusive ones using the approach of s.\ref{s.exin}.

\section{Racah matrices\label{s.rac}}

The Racah matrices are crucial elements for evaluating invariants of knots and links. There are two distinct sets of Racah matrices which appear in different approaches to studying knot polynomials: \underline{inclusive} and \underline{exclusive} Racah matrices.

\subsection{Inclusive Racah matrices\label{incrac}}

One of the approaches is originally due to N. Reshetikhin and V. Turaev \cite{RT}, hence the name Reshetikhin-Turaev (RT) approach.  According to this approach \cite{MMMI,RTmod}, in order to construct a knot/link invariant for the knot/link presented by a closed braid, one has to associate each crossing in this braid with a particular $\mathcal{R}$-matrix. If the studied object is a knot colored with a representation $R$, each strand in the braid carries the same representation $R$, all the $\mathcal{R}$-matrices act on the same tensor product $R\otimes R$, and the eigenvalues of all these $\mathcal{R}$-matrices are the same. $\mathcal{R}$-matrices acting on different pairs of strands are connected by rotation matrices. These rotation matrices appear to be Racah matrices \cite{MMMI}. Let us discuss in more details the three-strand case, which was mostly studied in our recent papers \cite{MMMI,MMMS21,MMMS31,IMMMec,Univ,BJLMMMS} and which we study in the present paper.

Within our approach, the object discussed is in fact not just the HOMFLY-PT polynomial but its character expansion. If one studies a three-strand knot in representation $R$, then the character expansion of its HOMFLY-PT polynomial includes all the representations $Q$ from the decomposition of the triple tensor cube of the representation $R$: $Q\vdash R^{\otimes 3}$. The HOMFLY-PT polynomial is then given by
\begin{equation}
H^K_R=\sum\limits_{Q\vdash R^{\otimes 3}} S^*_Q C_Q,
\end{equation}
where $S^*_Q$ are characters of representations $Q$ and $C_Q$ are coefficients constructed from the $\mathcal{R}$-matrices and Racah matrices. Thus, for each $Q$ for the three-strand knots, there exist two $\mathcal{R}$-matrices (the one corresponding to the crossing between upper pair of strands and the one corresponding to the lower pair) and they are related by rotation with a Racah matrix. One of the $\mathcal{R}$-matrices can be chosen diagonal, for the sake of definiteness, we choose the upper one as such, and the eigenvalues of the $\mathcal{R}$-matrix are described, up to a factor, by the eigenvalue of the second Casimir operator\footnote{
There is a simple formula for this eigenvalue
$$\varkappa_Y=1/2\sum_i Y_i (Y_i+1-2i)$$
which, in accordance with the Schur-Weyl duality, is associated with the value of character $\chi_Y([2])$ of the symmetric group  $S_{|Y|}$ in representation described by the Young diagram $Y$ on the cycle of length 2} $\varkappa_Y$:
\begin{equation}
\begin{array}{l}
\mathcal{R}_{Q;1}=diag(\epsilon_Y q^{\varkappa_Y}),\ Y\vdash R^{\otimes 2},\ Q\vdash Y\otimes R;
\\
\mathcal{R}_{Q;2}=U_Q \mathcal{R}_{Q;1} U^{\dagger}_Q;\ \ \ \ \ \ \ \ C_Q=Tr(\prod\limits_i \mathcal{R}_{Q;i}).
\end{array}
\end{equation}
Also needed for the definition of $\mathcal{R}$-matrices are $\epsilon_Y=\pm 1$, which describe the signs of these eigenvalues. At least for knots, the signs depend on if the representation $Y$ comes from symmetric ($+1$) or antisymmetric ($-1$) square of the initial representation $Q$, see \cite{MMMS21} for details. We discuss the issue of signs in more details in s.\ref{s.signs}.

We call the set of Racah matrices for \underline{all} representations $Q\vdash R^{\otimes 3}$ \underline{inclusive} Racah matrices. In traditional 6-j symbol notations, they can be written as
\begin{equation}
\label{UQ}
U_Q=U_{XY}\left[ \begin{array}{cc} R&R \\ R & Q \end{array} \right],\ X,Y\vdash R^{\otimes 2},\ Q\vdash X,Y\otimes R.
\end{equation}

If one discusses links rather than knots, the situation becomes slightly more complicated. Instead of one representation $R$ on all three strands, one could have three different representations $R_1$, $R_2$ and $R_3$. Consequently, there are two more diagonal $\mathcal{R}$-matrices for each $Q\vdash R_1\otimes R_2\otimes R_3$, i.e. there are three diagonal matrices in total: $\mathcal{R}_{12}$, $\mathcal{R}_{13}$ and $\mathcal{R}_{23}$. There are three independent Racah matrices as well: $U_{123}$, $U_{132}$ and $U_{213}$. Matrices corresponding to three remaining orderings coincide with the transposed matrices of the first three.

\begin{figure}[ht!]
\centering
\includegraphics[width=6 cm]{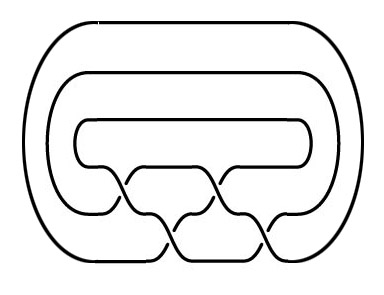}\ \ \ \ \ \includegraphics[width=6 cm]{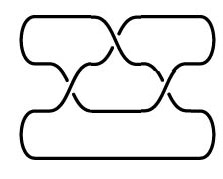}
\caption{Trefoil knot $3_1$ presented as a braid (on the left) and as a two-bridge knot(on the right).}
\label{knots}
\end{figure}

\subsection{Exclusive Racah matrices\label{exrac}}

Another approach to evaluating knot invariants is to study the two-bridge knots or their generalization, the arborescent knots \cite{Cau,arbor,Rama1}. Such knots have also representations different from the braid representation: the closure of the braid is made differently. From the point of view of representation theory, this leads to a different representation structure. E.g. if one studies the two-bridge knots, among all representations from the product of these four, only the trivial representation contributes, i.e. the four strands carry representations $R$, $R$, $\bar{R}$ and $\bar{R}$.  This means that, in the decomposition of the product of three representations out of four $R$, $R$, $\bar{R}$ and $\bar{R}$ discussed in s.\ref{incrac}, also only the single one contributes: the conjugate to the remaining fourth representation. In its turn, this implies that, in this case, there is no set of the inclusive Racah matrices $U_Q$ (\ref{UQ}), and only two matrices for each representation $R$ survive in the two-bridge case depending on the choice of the three representations out of four $R$, $R$, $\bar{R}$ and $\bar{R}$. Indeed, there are two $\mathcal{R}$-matrices for two bridge knots: one, which we call $T_R$ is associated with the crossing of two strands with either representations $R$ on the both strands or representations $\bar{R}$; another one, which we call $\bar{T}_R$ is associated with the crossing between different representations $R$ and $\bar{R}$. Consequently, in this case, there are, indeed, two Racah matrices, which we call \underline{exclusive}. These matrices relate $\mathcal{R}$-matrices acting on two strands in the middle and the ones, on two strands on the side (in the case of representations described by rectangular Young diagrams, both matrices acting on either upper pair of strands or on lower one coincide, while for other cases they typically differ). We denote $S$ the exclusive Racah matrices that intertwine the product $(R\otimes R)\otimes \bar{R}$ and the product $R\otimes (R\otimes \bar{R})$, and, $\bar{S}$ those intertwining the product $(R\otimes \bar{R})\otimes R$ and the product $R\otimes (\bar{R}\otimes R)$. From this definition, it immediately follows that $\bar{S}^{\dagger}=\bar{S}$. In the traditional 6-j symbol notation, these exclusive Racah matrices are written as
\begin{equation}
\begin{array}{l}
S_R=U_{XY}\left[ \begin{array}{cc} R&R \\ \bar{R} & R \end{array} \right],\
\begin{array}{l}
X\vdash R\otimes\bar{R},\ R\vdash R\otimes X, \\
Y\vdash R\otimes R,\ R\vdash Y\otimes \bar{R}.
\end{array}
\\
\bar{S}_R=U_{XY}\left[ \begin{array}{cc} R&\bar{R} \\ R & R \end{array} \right],\
\begin{array}{l}
X\vdash R\otimes\bar{R},\ R\vdash R\otimes X, \\
Y\vdash \bar{R}\otimes R,\ R\vdash Y\otimes R.
\end{array}
\end{array}
\end{equation}

For links, the picture is again more complicated. In this case, due to the construction, there can appear only two-component links colored by $R_1$ and $R_2$. Four strands now carry in any order representations $R_1$, $R_2$, $\bar{R}_1$ and $\bar{R}_2$. This provides four $\mathcal{R}$-matrices:
\begin{equation}
T_{12}=T_{\bar{1}\bar{2}},\ T_{1\bar{1}},\ T_{1\bar{2}}=T_{\bar{1}2},\ T_{2\bar{2}}.
\end{equation}
There are also three independent Racah matrices:
\begin{equation}
S_{1\bar{1}2}= S^{\dagger}_{\bar{1}2\bar{2}},\ S_{1\bar{1}\bar{2}}= S^{\dagger}_{\bar{1}\bar{2}2},\ S_{12\bar{1}}= S^{\dagger}_{2\bar{1}\bar{2}}.
\end{equation}
All other Racah matrices either coincide or are transposed of one of these three.

\begin{figure}[ht!]
\centering
\includegraphics[width=6 cm]{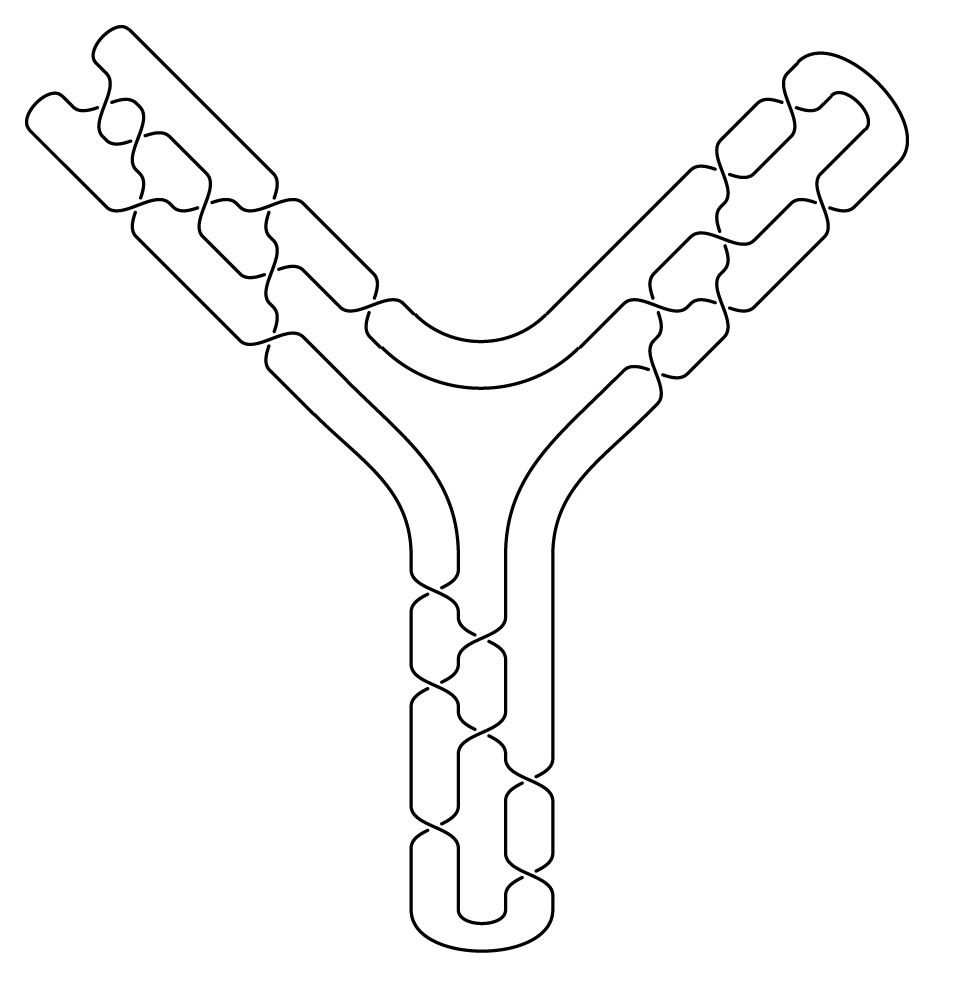}
\caption{Knot presented as an arborescent diagram.}
\label{knotarb}
\end{figure}

\section{Calculating Racah matrices\label{s.raccalc}}

The main approach to calculating Racah matrices uses representation theory of quantum groups and the highest weight vector calculations. This approach is described in detail in section \ref{hwc}.

Unfortunately, it is much harder to calculate exclusive Racah matrices using the same approach, since among other reasons the highest weight vectors of conjugate representations essentially depend on the group $SU(N)$ in contrast with non-conjugate representations. However, there is a ``trick'' \cite{MMMSint} which allows one to find the exclusive Racah matrices from the inclusive ones.

\subsection{Exclusive Racah through inclusive Racah\label{s.exin}}

This trick suggested in \cite{MMMSint} for knots is based on  studying a particular series of knots which are both three-strand and arborescent. One such example is given by the three-strand knots $(m,-1|\pm n,-1)$ in the notation of \cite{MMMI}, which, at the same time, are the Pretzel knots $Pr(m,n,\pm \bar{2})$ \cite{MMSpret}.

One can use the evolution method \cite{evo,MMSpret} to write down the answer for the whole series of knots. For the three-strand representation, one gets
\begin{equation}
H_R^{(m,-1|\pm n, -1)}=\sum\limits_{Y,Z\vdash R^{\otimes 2}} h_{YZ}\cdot \lambda_Y^m\lambda_Z^n.
\end{equation}
On the other side for the same series described as Pretzel knots \cite{MMSpret,Rama1}, one gets:
\begin{equation}
H_R^{Pr(m,n,\pm \bar{2})}=d^2_R \sum\limits_{\bar{X}\vdash R\otimes\bar{R}}
\cfrac{(ST^mS^{\dagger})_{\emptyset \bar{X}}(ST^nS^{\dagger})_{\emptyset \bar{X}}(\bar{S}\bar{T}^{\pm 2}\bar{S})_{\emptyset \bar{X}}}{\bar{S}_{\emptyset \bar{X}}}=
\sum\limits_{{\bar{X}\vdash R\otimes\bar{R}}\atop{Y,Z\vdash R\otimes R}}\sqrt{d_Y d_Z} K_{\bar{X}}S_{\bar{X}Y}S_{\bar{X}Z}\cdot\lambda_Y^m\lambda_Z^n,
\end{equation}
where $\lambda_Y$ and $\lambda_Z$ are the eigenvalues of $\mathcal{R}$-matrices in the three strand case, they form the diagonal matrix $T$ from the point of view of the Pretzel knots (they coincide, because they are associated with the crossing with the same representations).

The equality between these two formulae can be rewritten as
\begin{equation}
\sum\limits_{\bar{X}} \left(K_{\bar{X}} (\bar{S}\bar{T}^{\pm 2}\bar{S})_{\emptyset\bar{X}}\right)S_{\bar{X}Y}S_{\bar{X}Z}=\cfrac{h_{YZ}}{\sqrt{d_Y d_Z}}.
\end{equation}
Thus, the matrix $S$ can be found as the matrix that diagonalizes the matrix at the r.h.s.

In the case of links, the situation is as always more difficult. One can use the same series to study links as well. To this end, $m$ and $n$ should be even (we will just put $2m$ and $2n$ instead). Then, the three-strand braid $(2m,-1|2n,\pm 1)$ is a three component link. If we color it with representations $R_1$, $R_2$ and $R_3$, the resulting polynomial is
\begin{equation}
H_{123}^{(2m,-1|2n,\pm 1)}=\sum\limits_{Q\vdash R_1\otimes R_2\otimes R_3}S^*_Q \text{Trace}\left(
\mathcal{R}^{2m}_{12} U_{123} \mathcal{R}^{-1}_{23} U_{132}^{\dagger} \mathcal{R}^{2n}_{13} U_{132} \mathcal{R}^{-1}_{23} U_{123}^{\dagger}
\right)_Q=\sum\limits_{{Y\vdash R_1\otimes R_2}\atop{Z\vdash R_1\otimes R_3}} h_{YZ}\lambda_Y^{2m}\lambda_Z^{2n}.
\end{equation}
At the same time, evaluating the same links from the Pretzel representation gives the following expression
\begin{equation}
\begin{array}{r}
H_{123}^{Pr(2m,2n,\pm \bar{2})}=d^2_{R_1} \sum\limits_{\bar{X}\vdash R_1\otimes\bar{R_1}}
\cfrac{(S_{1\bar{1}\bar{2}}T_{12}^{2m}S_{1\bar{1}\bar{2}}^{\dagger})_{\emptyset \bar{X}}(S_{1\bar{1}\bar{3}}T_{13}^{2n}S_{1\bar{1}\bar{3}}^{\dagger})_{\emptyset \bar{X}} (S_{2\bar{2}3}T_{2\bar{3}}^{\pm 2}S_{2\bar{2}3}^{\dagger})_{\emptyset \bar{X}}}{\bar{S}_{\emptyset \bar{X}}}=
\\ \\
=\sum\limits_{{{\bar{X}\vdash R_1\otimes\bar{R_1}}\atop{Y\vdash R_1\otimes R_2}}\atop{Z\vdash R_1\otimes R_3}}\sqrt{d_Y d_Z} K_{\bar{X}}S_{1\bar{1}\bar{2}|\bar{X}Y}S_{1\bar{1}\bar{3}|\bar{X}Z}\cdot\lambda_Y^{2m}\lambda_Z^{2n}.
\end{array}
\end{equation}
Here we assume that, in the case of symmetric representations, $R_1$ is the smallest representation and dimensions are defined by this smallest representations. This is due to the fact that the set of representations $\bar{X}$ is defined by the smallest of the representations between $R_1$, $R_2$ and $R_3$. If representations are not symmetric then the set of $\bar{X}$ and dimensions are defined by the intersection of sets of representations $R_1\otimes \bar{R}_1$, $R_2\otimes \bar{R}_2$ and $R_3\otimes \bar{R}_3$.
Comparing these relations, one gets the equation
\begin{equation}
K_{\bar{X}}S_{1\bar{1}\bar{2}|\bar{X}Y}\lambda_Y^{2m}S_{1\bar{1}\bar{3}|\bar{X}Z}\lambda_Z^{2n}=\cfrac{h_{YZ}}{\sqrt{d_Y d_Z}}.
\end{equation}
Repeating this procedure for other placements of colors, one can get three equations for two matrices each. Solving them will provide the three exclusive Racah matrices $S_{1\bar{1}\bar{2}}$, $S_{1\bar{1}\bar{3}}$ and $S_{2\bar{2}\bar{3}}$. Similar equations can be constructed for other matrices $S$.

We provide a non-trivial essentially link-related example of calculating the exclusive Racah matrices in s.\ref{s.exex}.

\subsection{Symmetries of Racah matrices}

The Racah matrices possess some symmetry properties, which allow one to greatly reduce the number of Racah coefficients that require direct computation. The main property is general for all the constructions in quantum groups: the substitution $q\rightarrow -q^{-1}$ corresponds to the transposition of all the Young diagrams associated with the representations. This means that if one has calculated the Racah matrix $U_{XY}\left[ \begin{array}{cc} R&R \\ R & Q \end{array} \right]$, it also has provided the answer for $U_{X^TY^T}\left[ \begin{array}{cc} R^T&R^T \\ R^T & Q^T \end{array} \right]$.

Another important symmetry property of the Racah matrix is that, for the inclusive Racah matrices, one can reverse the order of multiplication of all representations. This property was already mentioned earlier, basically it means that the Racah matrix appearing in the product $R_1\otimes R_2\otimes R_3$ is inverse to the one appearing in the product $R_3\otimes R_2\otimes R_1$.

\subsection{The highest weight calculus \label{hwc}}

Here we repeat the method to calculate the Racah matrices used in our previous papers \cite{MMMI,MMMS21,MMMS31,IMMMec,Univ,BJLMMMS}.

The Racah matrix $U$ is a transformation matrix from one orthonormal basis (I) to another one (II), which are defined as follows:
\begin{equation}
U_{XY}\left[ \begin{array}{cc} R_1&R_2\\R_3&R_4 \end{array} \right]: \left( \underbrace{R_1 \otimes R_2}_X \right) \otimes R_3 \xrightarrow{ \ I \ } Q \ \ \rightarrow \ \  R_1 \otimes \left( \underbrace{R_2 \otimes R_3}_Y \right) \xrightarrow{ \ II \ } Q.
\end{equation}
For example, in the case $R_1=[3],R_2=[2,1],R_3=[1,1,1]$ and $Q$ is arbitrary, most of the Racah matrices are equal to the identity matrix except for few ones. To find $Q$ giving non-trivial contributions, one can use the Littlewood-Richardson rule:
\begin{equation}
\begin{array}{ccl}\label{[3][21][111]dec}
\chi_{R_1} \cdot \chi_{R_2} &=& \sum_Q C_{R_1,R_2}^{Q} \cdot \chi_Q \\
\chi_{[3]} \cdot \chi_{[2,1]} \cdot \chi_{[1,1,1]} &=& \chi_{[6,2,1]}+\chi_{[6,1,1,1]}+\chi_{[5,3,1]}+3\,\chi_{[5,2,1,1]}+2\,\chi_{[5,1,1,1,1]}+\chi_{[4,3,2]}+2\,\chi_{[4,3,1,1]} \\ &+& 2\,\chi_{[4,2,2,1]}+ 3\,\chi_{[4,2,1,1,1]}+\chi_{[4,1,1,1,1,1]} + \chi_{[3,3,2,1]}+\chi_{[3,3,1,1,1]}+\chi_{[3,2,2,1,1]} \\ &+&\chi_{[3,2,1,1,1,1]},
\end{array}
\end{equation}
where $\chi_R$ is the character of the irreducible representation, which is the Schur function in the case of $SU(N)$, while $R$'s in this case are labelled by the Young diagrams. From now on, we identify the representations with the Young diagrams.

The coefficients $C_{R_1,R_2}^{Q}$ count how many times the irreducible representation $Q$ appears in the decomposition, therefore they determine the size of the corresponding Racah matrix. Decomposition (\ref{[3][21][111]dec}) shows us that there are two matrices of size $3\times3$, three matrices of size $2\times2$ and ten trivial ``matrices'' of size $1\times1$.

We calculate the Racah matrix using its definition, i.e. as a transformation matrix from the orthonormal basis (I) to the orthonormal basis (II). To this end, we construct the highest weight vectors in the basis (I) for each representation $Q$ from (\ref{[3][21][111]dec}) and same in the basis (II). To proceed, we need to describe manifestly the action of lowering and raising operators $T_k^{\pm}$ on representations of $U_q(sl_N)$.

To this end, we use the Schur-Weyl duality and, first, realize the representation of $U_q(sl_N)$ in the space of tensors. With each representation labeled by the Young diagram $Y = \{ Y_1\geq Y_2\geq\ldots\geq Y_l>0 \}$, we associate the following tensor with all possible permutations of indices:
\begin{equation}\label{elY}
V_{i_1,\dots,i_{Y_1},j_1,\dots,j_{Y_2},k_1,\dots,k_{Y_3},\ldots}, \text{where}\
i_1=\dots=i_{Y_1}=0, \ j_1=\dots=j_{Y_2}=1, \ k_1=\dots=k_{Y_3}=2, \ldots .
\end{equation}
In other words, the number of zeros is equal to $Y_1$, the number of units is equal to $Y_2$, the number of deuces is equal to $Y_3$ and so on. Thus, every vector of the representation $Y$ can be written as a linear combination of elements (\ref{elY}).

Second, let us define the action of lowering and raising operators $T_k^{\pm}$. It is clear that for 1-tensors they act as follows:
\begin{equation}
\begin{array}{l}
T_k^{+}: \left\{ \begin{array}{lcl}V_{k-1} & \longrightarrow & V_k, \\ V_i &\longrightarrow & 0 (i\neq k-1), \end{array}\right.\\
T_k^{-}: \left\{ \begin{array}{lcl}V_k & \longrightarrow & V_{k-1}, \\ V_i & \longrightarrow & 0 (i\neq k). \end{array}\right.
\end{array}
\end{equation}

To extend this action to higher rank tensors, one needs a uniquely defined comultiplication $\Delta$ on $U_q(sl_N)$:
\begin{equation}
\begin{array}{lcl}
\Delta(E_i) &=& 1\otimes E_i + E_i\otimes q^{H_i}, \\
\Delta(F_i) &=& F_i\otimes 1 + q^{-H_i}\otimes F_i, \\
\Delta(q^{H_i}) &=& q^{H_i}\otimes q^{H_i},
\end{array}
\end{equation}
where $E_i, F_i, H_i(1\leq i \leq N-1)$ are generators of $U_q(sl_N)$.

Since $Y$ is a representation of $U_q(sl_N)$, it means there is a given algebra homomorphism between them. If we denote $T_k^{+}$ and $T_k^-$ to be the images of $E_k$ and $F_k$ respectively, one gets for 2-tensors:
\begin{equation}
\begin{array}{ll}
T_k^{+}: V_{i,j} \longrightarrow & \delta_{k-1,j}V_{i,j+1} + \delta_{k-1,i} q^{H_k}(V_j) V_{i+1,j},  \\
T_k^{-}: V_{i,j} \longrightarrow & \delta_{k,j}q^{-H_k}(V_i) V_{i,j-1} + \delta_{k,i} V_{i-1,j}.
\end{array}
\end{equation}
Here $\delta_{i,j}$ is the usual Kronecker symbol, and the notation $q^{H_k}(V_i)$ is defined as follows:
\begin{equation}
q^{H_k}: \left\{ \begin{array}{lcl}V_{k-1} & \longrightarrow & q, \\ V_k &\longrightarrow & q^{-1}, \\ V_i &\longrightarrow & 1 \ (i\neq k-1 \ or \ k). \end{array}\right.
\end{equation}
Actually it coincides with the action of $H_k$ in a representation space of $U_q(sl_N)$:
\begin{equation}
H_k: \left\{ \begin{array}{lcl}v_{k-1} & \longrightarrow & v_{k-1}, \\ v_k &\longrightarrow & -v_k, \\ v_i &\longrightarrow & 0 \ (i\neq k-1 \ or \ k). \end{array}\right.
\end{equation}

Since $\Delta$ is co-associative, it is easy to extend $T_k^{\pm}$ actions to any rank tensors.

\subsection{Example 1. $[1]\otimes[1]\otimes[2]$ \label{ex112}}

The decomposition in this case takes the form
\begin{equation}
[1]\otimes[1]\otimes[2] = [4] \oplus 2[3,1] \oplus [2,2] \oplus [2,1,1].
\end{equation}
The Racah matrices for $[1]\otimes[1]\otimes[2]\to \{[4], [2,2], [2,1,1]\}$ are just equal to 1. The only non-trivial matrix is for $[1]\otimes[1]\otimes[2]\to [3,1]$, and it relates the bases
\begin{equation}
([1]\otimes [1])\otimes [2]=([2]\oplus[1,1])\otimes[2]
\end{equation}
and
\begin{equation}
[1]\otimes ([1]\otimes [2])=[1]\otimes([3]\oplus[2,1]).
\end{equation}
The corresponding highest weight vectors are
\begin{equation}\label{u1}
\begin{array}{l}
u_1 = \dfrac{1}{\sqrt{q^6+q^4+q^2+1}}\left(q^3v_{0 0 1 0}+q^2v_{0 0 0 1}-qv_{1 0 0 0}-v_{0 1 0 0}\right), \\
u_2 = \dfrac{1}{\sqrt{q^2+1}}\left(qv_{0 1 0 0}-v_{1 0 0 0}\right), \\
u^{'}_1 = \dfrac{1}{\sqrt{(q^6+q^4+q^2+1)(q^4+q^2+1)}}\left(q^5v_{0 1 0 0}+q^4v_{0 0 1 0}-q^4v_{1 0 0 0}+q^3v_{0 0 0 1}-q^2v_{1 0 0 0}-v_{1 0 0 0}\right), \\
u^{'}_2 = \dfrac{1}{\sqrt{(q^4+q^2+1)(q^2+1)}}\left(q^3v_{0 0 1 0}+q^2v_{0 0 0 1}-q^2v_{0 1 0 0}-v_{0 1 0 0}\right).
\end{array}
\end{equation}

In order to demonstrate how to find the highest weight vectors, let us obtain $u_1$ as an example. From the definition, it is obvious that $v_{00}$ is just the highest weight vector in representation $[2]$. The action of $T_1^{+}$ on $v_{00}$ gives $v_{01}+qv_{10}$. Since $u_1$ is the highest weight vector in $[3,1]\vdash[2]\otimes[2]$, it should be a linear combination of two vectors $w_1$ and $w_2$, obtained by the action of $T_1^{+}$ on the first two indices of $v_{0000}$ and on the second two indices of $v_{0000}$ correspondingly:
$u_1=\alpha w_1+\beta w_2$, $w_1=v_{0100}+qv_{1000}$, $w_2=v_{0001}+qv_{0010}$.
To determine $\alpha$ and $\beta$, one requires that all lowering operators $T_k^-$ cancel on arbitrary vector from this space:
\begin{equation}
T_k^-\left( \alpha \left( v_{0100}+qv_{1000} \right)+\beta \left( v_{0001}+qv_{0010}\right)\right) = 0 \Rightarrow \alpha=-c,\beta=q^2\cdot c,
\end{equation}
where $c$ is an arbitrary constant.

By definition, the Racah matrix is a transformation matrix from one orthonormal basis to another one, hence, all the highest weight vectors have unit norms. This determines $c$ uniquely up to a sign:
\begin{equation}
c = \pm\dfrac{1}{\sqrt{q^6+q^4+q^2+1}}.
\label{c31}
\end{equation}
In the answer above, we choose the sign to be '+'. This gives us the highest weight vector $u_1$ in (\ref{u1}).

It is immediate now to obtain the Racah matrix:
\begin{equation}
U\left[ \begin{array}{cc} [1]&[1] \\ \text{[2]} & [3,1] \end{array} \right] = \left( \begin{array}{cc} \dfrac{1}{\sqrt{[3]}}&\dfrac{\sqrt{[4]}}{\sqrt{[2][3]}} \\ \dfrac{\sqrt{[4]}}{\sqrt{[2][3]}} & -\dfrac{1}{\sqrt{[3]}} \end{array} \right).
\label{u112ans}
\end{equation}

\subsection{Example 2. $[1]\otimes[2]\otimes[1]$}

The decomposition in this case takes the form
\begin{equation}
[1]\otimes[2]\otimes[1] = [4] \oplus 2[3,1] \oplus [2,2] \oplus [2,1,1]
\end{equation}

The Racah matrices for $[1]\otimes[2]\otimes[1]\to\{[4], [2,2], [2,1,1]\}$ are just equal to 1. The only non-trivial matrix is for $[1]\otimes[2]\otimes[1]\to [3,1]$ and it relates the bases
\begin{equation}
([1]\otimes [2])\otimes [1]=([3]\oplus[2,1])\otimes[1]
\end{equation}
and
\begin{equation}
[1]\otimes ([2]\otimes [1])=[1]\otimes([3]\oplus[2,1]).
\end{equation}
The corresponding highest weight vectors are
\begin{equation}
\begin{array}{l}
u_1 =\dfrac{1}{\sqrt{(1+q^2)(1+q^4)(1+q^2+q^4)}}\left(q(1+q^2+q^4)v_{0 0 0 1}-v_{0 0 1 0}-qv_{0 1 0 0}-q^2v_{1 0 0 0}\right),
\\
u_2 =\dfrac{1}{\sqrt{(1+q^2)(1+q^2+q^4)}}\left(q^2v_{0 0 1 0}+q^3v_{0 1 0 0}-(q^2+1)v_{1 0 0 0}\right),
\\
u^{'}_1 =\dfrac{1}{\sqrt{(1+q^2)(1+q^4)(1+q^2+q^4)}}\left(q^3v_{0 0 0 1}+q^4v_{0 0 1 0}+q^5v_{0 1 0 0}-(1+q^2+q^4)v_{1 0 0 0}\right),
\\
u^{'}_2 =\dfrac{1}{\sqrt{(1+q^2)(1+q^2+q^4)}}\left(-q(q^2+1)v_{0 0 0 1}+v_{0 0 1 0}+qv_{0 1 0 0}\right).
\end{array}
\end{equation}

It is now immediate to calculate the Racah matrix:
\begin{equation}
U\left[ \begin{array}{cc} [1]&[2] \\ \text{[1]} & [3,1] \end{array} \right] = \left( \begin{array}{cc} \dfrac{1}{[3]}&\dfrac{\sqrt{[4][2]}}{[3]} \\ \dfrac{\sqrt{[4][2]}}{[3]} & -\dfrac{1}{[3]} \end{array} \right).
\end{equation}

\subsection{Example 3. $[1]\otimes[2]\otimes[2,1]$}

The decomposition in this case takes the form
\begin{equation}
[1]\otimes[2]\otimes[2,1] = [5,1]\oplus 2[4,2] \oplus 2[4,1,1] \oplus [3,3] \oplus 3[3,2,1] \oplus [3,1,1,1] \oplus [2,2,2] \oplus [2,2,1,1].
\end{equation}
The Racah matrices for $[1]\otimes[2]\otimes[2,1]\to\{[5,1], [3,3], [3,1,1,1], [2,2,2], [2,2,1,1]\}$ are just equal to 1. The only non-trivial matrices are for $[1]\otimes[2]\otimes[2,1]\to\{[4,2], [4,1,1], [3,2,1]\}$ and they relate the bases
\begin{equation}
([1]\otimes [2])\otimes [2,1]=([3]\oplus[2,1])\otimes[2,1]
\end{equation}
and
\begin{equation}
[1]\otimes ([2]\otimes [2,1])=[1]\otimes([4,1]\oplus[3,2]\oplus[3,1,1]\oplus[2,2,1]).
\end{equation}
The corresponding highest weight vectors of representations $[4,2]$ are
\begin{equation}
\begin{array}{ll}
    u_1 =&\dfrac{1}{(q^2+1)\sqrt{(1+q^4)(1+q^2+q^4)}}(q(q^4+q^2+1)(qv_{000011}-v_{000101})-q^3v_{100010}+q^2v_{100100}\\
         &+qv_{010100}-q^2v_{010010}+v_{001100}-qv_{001010}),\\
    u_2 =&\dfrac{1}{(q^2+1)\sqrt{(1+q^2+q^4)}}(q^4v_{010010}-q^3v_{010100}+q^3v_{010100}\\
         &-q^2v_{001100}-q(q^2+1)v_{100010}+(q^2+1)v_{100100}),\\
u^{'}_1 =&\dfrac{1}{(q^2+1)\sqrt{(1+q^4)(1+q^2+q^4)}}(-(1+q^2+q^4)(qv_{100010}-v_{100100})+q^4v_{000011}+q^5v_{001010}\\
         &+q^6v_{010010}-q^3v_{000101}-q^4v_{001100}-q^5v_{010100}),\\
u^{'}_2 =&\dfrac{1}{(q^2+1)\sqrt{(1+q^2+q^4)}}\left(qv_{001010}+q^2v_{010010}-v_{001100}-qv_{010100}-q(1+q^2)(qv_{000011}-v_{000101})\right).
\end{array}
\end{equation}
It is now immediate to find out the Racah matrix for $[4,2]$:
\begin{equation}
U\left[ \begin{array}{cc} [1]&[2] \\ \text{[2,1]} & [4,2] \end{array} \right] = \left( \begin{array}{cc} \dfrac{1}{\sqrt{[3]}}&-\dfrac{\sqrt{[4][2]}}{[3]} \\ \dfrac{\sqrt{[4][2]}}{[3]} & \dfrac{1}{\sqrt{[3]}} \end{array} \right).
\end{equation}
The highest weight vectors in representations $[3,2,1]$ and $[4,1,1]$ are too cumbersome to be written down here. They can be found in \cite{knotebook}. The Racah matrix for $[3,2,1]$ is
\begin{equation}
U\left[ \begin{array}{cc} [1]&[2] \\ \text{[2,1]} & [3,2,1] \end{array} \right] = \left( \begin{array}{ccc} -\dfrac{1}{\sqrt{[2][4]}}&\dfrac{1}{\sqrt{[2]}}&\dfrac{\sqrt{[5]}}{\sqrt{[2][4]}} \\ \dfrac{\sqrt{[5]}}{\sqrt{[4]!}} & -\dfrac{\sqrt{[5]}}{[2]\sqrt{[3]}} & \dfrac{\sqrt{[3]}}{\sqrt{[2][4]}} \\ \dfrac{\sqrt{[4]}}{\sqrt{[2][3]}} & \dfrac{1}{\sqrt{[3]}} &0 \end{array} \right),
\end{equation}
and for $[4,1,1]$ is
\begin{equation}
U\left[ \begin{array}{cc} [1]&[2] \\ \text{[2,1]} & [4,1,1] \end{array} \right] = \left( \begin{array}{cc} \dfrac{1}{\sqrt{[5]}}&-\dfrac{\sqrt{[4][2]}}{\sqrt{[3][5]}} \\ \dfrac{\sqrt{[4][2]}}{\sqrt{[3][5]}} & \dfrac{1}{\sqrt{[5]}} \end{array} \right).
\end{equation}

\section{List of Racah matrices\label{s.raclist}}

We have calculated all the inclusive Racah matrices for the representations described by Young diagrams with 3 boxes. In fact, the case of  $U_{XY}\left[ \begin{array}{cc} [2,1]&[2,1] \\ \text{[2,1]} & Q \end{array} \right]$ has been calculated earlier \cite{MMMS21} (see also \cite{GJ} for the exclusive Racah matrices in the $[2,1]$ case), hence, here we calculate the following Racah matrices:
{\small
\begin{equation}\label{Rac}
\begin{array}{l}
U_{XY}\left[ \begin{array}{cc} [3]&[2,1] \\ \text{[3]} & Q \end{array} \right],  \
U_{XY}\left[ \begin{array}{cc} [3]&[1,1,1] \\ \text{[3]} & Q \end{array} \right],   \
U_{XY}\left[ \begin{array}{cc} [2,1]&[3] \\ \text{[3]} & Q \end{array} \right],   \
U_{XY}\left[ \begin{array}{cc} [2,1]&[2,1] \\ \text{[3]} & Q \end{array} \right],   \\
U_{XY}\left[ \begin{array}{cc} [2,1]&[1,1,1] \\ \text{[3]} & Q \end{array} \right] \
U_{XY}\left[ \begin{array}{cc} [1,1,1]&[3] \\ \text{[3]} & Q \end{array} \right],   \
U_{XY}\left[ \begin{array}{cc} [1,1,1]&[2,1] \\ \text{[3]} & Q \end{array} \right],   \
U_{XY}\left[ \begin{array}{cc} [1,1,1]&[1,1,1] \\ \text{[3]} & Q \end{array} \right],   \\
U_{XY}\left[ \begin{array}{cc} [3]&[3] \\ \text{[2,1]} & Q \end{array} \right],   \
U_{XY}\left[ \begin{array}{cc} [3]&[2,1] \\ \text{[2,1]} & Q \end{array} \right]  \
U_{XY}\left[ \begin{array}{cc} [3]&[1,1,1] \\ \text{[2,1]} & Q \end{array} \right],   \
U_{XY}\left[ \begin{array}{cc} [2,1]&[3] \\ \text{[2,1]} & Q \end{array} \right],   \\
U_{XY}\left[ \begin{array}{cc} [2,1]&[1,1,1] \\ \text{[2,1]} & Q \end{array} \right],   \
U_{XY}\left[ \begin{array}{cc} [1,1,1]&[3] \\ \text{[2,1]} & Q \end{array} \right],   \
U_{XY}\left[ \begin{array}{cc} [1,1,1]&[2,1] \\ \text{[2,1]} & Q \end{array} \right],  \
U_{XY}\left[ \begin{array}{cc} [1,1,1]&[1,1,1] \\ \text{[2,1]} & Q \end{array} \right],    \\
U_{XY}\left[ \begin{array}{cc} [3]&[3] \\ \text{[1,1,1]} & Q \end{array} \right],   \
U_{XY}\left[ \begin{array}{cc} [3]&[2,1] \\ \text{[1,1,1]} & Q \end{array} \right],   \
U_{XY}\left[ \begin{array}{cc} [3]&[1,1,1] \\ \text{[1,1,1]} & Q \end{array} \right],   \
U_{XY}\left[ \begin{array}{cc} [2,1]&[3] \\ \text{[1,1,1]} & Q \end{array} \right],  \\
U_{XY}\left[ \begin{array}{cc} [2,1]&[2,1] \\ \text{[1,1,1]} & Q \end{array} \right],   \
U_{XY}\left[ \begin{array}{cc} [2,1]&[1,1,1] \\ \text{[1,1,1]} & Q \end{array} \right],    \
U_{XY}\left[ \begin{array}{cc} [1,1,1]&[3] \\ \text{[1,1,1]} & Q \end{array} \right],  \
U_{XY}\left[ \begin{array}{cc} [1,1,1]&[2,1] \\ \text{[1,1,1]} & Q \end{array} \right].
\end{array}
\end{equation}
}
Decompositions of the products of the representations involved in these Racah matrices are
\begin{equation}
\begin{array}{lcl}
[3]\otimes [3] &=& [6]\oplus [5, 1]\oplus [4, 2]\oplus [3, 3], \\ {}
[3]\otimes [2,1] &=& [5,1]\oplus [4,2] \oplus [4,1,1] \oplus [3,2,1], \\ {}
[3]\otimes [1,1,1] &=& [4,1,1]\oplus [3,1,1,1], \\ {}
[2,1]\otimes [2,1] &=& [4,2] \oplus [4,1,1] \oplus [3,3] \oplus 2[3,2,1] \oplus [3,1,1,1] \oplus [2,2,2] \oplus [2,2,1,1], \\ {}
[2,1]\otimes [1,1,1] &=& [3,2,1]\oplus [3,1,1,1] \oplus [2,2,1,1]\oplus [2,1,1,1,1], \\ {}
[1,1,1]\otimes [1,1,1] &=& [2,2,2]\oplus [2,2,1,1] \oplus [2,1,1,1]\oplus [1,1,1,1,1,1].
\end{array}
\end{equation}
The maximal size of the Racah mixing matrices will be $5\times 5$: for $U_{XY}\left[ \begin{array}{cc} [2,1]&[2,1] \\ \text{[3]} & [5,2,1,1] \end{array} \right]$ and $U_{XY}\left[ \begin{array}{cc} [2,1]&[2,1] \\ \text{[3]} & [5,3,1] \end{array} \right]$. All representations come without multiplicities, except for the $[2,1]$ case.

As an example, consider the product
\begin{equation}
\begin{array}{lcl}
[3]\otimes [3]\otimes[2,1] &=&
[8, 1]+2[7, 2]+2[7, 1, 1]+2[6, 3]+4[6, 2, 1]+[6, 1, 1, 1]+2[5, 4]+4[5, 3, 1]+ \\
&+&2[5, 2, 2]+2[5, 2, 1, 1]+2[4, 4, 1]+2[4, 3, 2]+2[4, 3, 1, 1]+[4, 2, 2, 1]+[3, 3, 2, 1].
\end{array}
\end{equation}
Hence, the inclusive Racah matrices $U_{XY}\left[ \begin{array}{cc} [3]&[3] \\ \text{[2,1]} & Q \end{array} \right]$ form the collection

{\scriptsize
\begin{equation*}
\begin{array}{|c|p{14cm}|c|}
\hline
&&\text{number of}\\
\text{matrix size} & \hspace{6.2cm} Q &  \\
&&\text{matrices}\\ \hline && \\
1 & [8,1],[6,1,1,1],[4,2,2,1],[3,3,2,1] & 4 \\
&&\\ \hline && \\
2 & [7,2],[7,1,1],[6,3],[5,4],[5,2,2],[5,2,1,1],[4,4,1],[4,3,2],[4,3,1,1] & 9 \\
&&\\ \hline && \\
3 &  & 0 \\
&&\\ \hline && \\
4 & [5,3,1],[6,2,1] & 2 \\
&&\\ \hline && \\
5 &  & 0 \\
&&\\ \hline
 \end{array}
\end{equation*}
}

All the Racah matrices (\ref{Rac}) were calculated with the help of the highest weight method. The results are available online at \cite{knotebook}. Here we list only two Racah matrices as an illustration:
\begin{equation}
U_{XY}\left[ \begin{array}{cc} [3]&[3] \\ \text{[2,1]} & [5,4] \end{array} \right]=
\left(\begin{array}{rr}
\sqrt{\cfrac{[3]}{[2][4]}} & \sqrt{\cfrac{[5]}{[2][4]}} \\
& \\
\sqrt{\cfrac{[5]}{[2][4]}} & -\sqrt{\cfrac{[3]}{[2][4]}}
\end{array}\right),
\end{equation}
\begin{equation}
U_{XY}\left[ \begin{array}{cc} [3]&[2] \\ \text{[2]} & [5,2] \end{array} \right]=
\left(\begin{array}{rrr}
\cfrac{[2]}{[4]\sqrt{[5]}} & \cfrac{\sqrt{[2][3][6]}}{[4]\sqrt{[5]}} & \sqrt{\cfrac{[6]}{[3][4]}} \\
&& \\
\cfrac{[2]^2}{[3]}\sqrt{\cfrac{[6]}{[4][5]}} & \cfrac{[8]}{[4]}\sqrt{\cfrac{[2]}{[3][4][5]}} & -\cfrac{1}{\sqrt{[3]}} \\
&& \\
\cfrac{\sqrt{[2][5][6]}}{[3][4]} &-\cfrac{[2]}{[4]}\sqrt{\cfrac{[5]}{[3]}}& \sqrt{\cfrac{[2]}{[3][4]}}
\end{array}\right).
\end{equation}

\section{Signs of the $\mathcal{R}$-matrix eigenvalues\label{s.signs}}

Now we will discuss the issue of signs of the $\mathcal{R}$-matrix eigenvalues. While their absolute values were already discussed in s.\ref{incrac},  the sign issue is more complicated. Let us first discuss the knot case.

The $\mathcal{R}$-matrices for knots act on the tensor square of the representation associated with the strands:
\begin{equation}
\mathcal{R}:\ \ T\otimes T \rightarrow T\otimes T.
\end{equation}
The sign of eigenvalue depends on whether the representation under consideration, $Q\vdash T\otimes T$ in the classical case $q=1$ is symmetric or antisymmetric under permutation of two representations $T$. It is same to say that the representation $Q$ belongs to the symmetric or antisymmetric squares of representation $T$, see also \cite{MMMS21,Rama2}.

For links, the representations on which the $\mathcal{R}$-matrix acts may differ:
\begin{equation}
\mathcal{R}:\ \ T_1\otimes T_2 \rightarrow T_2\otimes T_1.
\end{equation}
Let us look at the structure of representations. As can be seen from the examples in the previous section, e.g. (\ref{c31}) the highest weight vectors and consequently the whole representation vectors are defined up to a sign. For the knot case, it makes no difference since the change of signs of representation vectors changes for the both spaces which the $\mathcal{R} $ acts on. However, for the link case, the two spaces are different, and the signs of the $\mathcal{R}$-matrix \underline{cannot be defined} in a unique way. They in fact depend on the choice of the signs of the representations.

This leads to another interesting property. Since the Racah matrices come from the products of three representations which are again defined up to a sign, it means that, by changing the signs of the basis vectors on one or another side of the Racah-matrix one can change the sign of a column or a row in the Racah matrix. For knots, this would not change the diagonal (or, in fact, any) $\mathcal{R}$-matrix, since its eigenvalues are uniquely defined. However, for links, the signs of the eigenvalues also depend on the signs of representation vectors. This means that one should define the signs in both the $\mathcal{R}$-matrices and the Racah matrices at once.

\section{Eigenvalue conjecture and Racah matrices\label{s.eig}}

In \cite{IMMMec}, an eigenvalue conjecture for the Racah matrices was suggested. This conjecture states that the inclusive Racah matrices in the three-strand knots depend only on normalized eigenvalues of the corresponding $\mathcal{R}$-matrix. The idea behind this conjecture, besides an experimental evidence, is quite simple. The defining property of the $\mathcal{R}$-matrices is that it should satisfy the Yang-Baxter equation (it is, in fact, nothing but the third Reidemeister move in knot theory), which, for the three-strand braid, looks like
\begin{equation}
\mathcal{R}_1\mathcal{R}_2\mathcal{R}_1=\mathcal{R}_2\mathcal{R}_1\mathcal{R}_2.
\end{equation}
If one also expresses here $\mathcal{R}_2$ through the Racah matrices
\begin{equation}
\mathcal{R}_2=U\mathcal{R}_1 U^{\dagger},
\end{equation}
an equation which relates the Racah coefficients with the $\mathcal{R}$-matrix appears:
\begin{equation}
\mathcal{R}_1U\mathcal{R}_1 U^{\dagger}\mathcal{R}_1=U\mathcal{R}_1 U^{\dagger}\mathcal{R}_1 U\mathcal{R}_1 U^{\dagger}.
\label{YBU}
\end{equation}
Together with the fact that the Racah matrix is, in fact, a rotation matrix (unitary or orthogonal in the real case), this equation can be solved for matrices of the sizes at least up to $6\times 6$ \cite{IMMMec,Univ} (see also \cite{Wentzl}). The solutions are unique for the $\mathcal{R}$-matrix of a generic form \footnote{
Of course, there are degenerate solutions which correspond to the Racah matrices of smaller sizes, e.g. the matrix of the size $6\times 6$ can be ``block-diagonal'' and be divided into two or more matrices of smaller sizes. These matrices would definitely solve the Yang-Baxter equation too, and their smaller compartments would satisfy their own smaller version of the eigenvalue conjecture. Hence, the solution is unique if one considers the Racah matrix which ``mixes'' \underline{all} the eigenvalues (representations) in $\mathcal{R}$-matrix. }
up to inessential choice of signs and are given in their full form in \cite{IMMMec,Univ,Wentzl}. Here we will only provide the answer for the Racah matrix of the size $2\times 2$ as an example.

If the $\mathcal{R}$-matrix is
\begin{equation}
\mathcal{R}_1=\left(\begin{array}{rr}\lambda_1 & \\ & \lambda_2\end{array}\right)=\sqrt{|\lambda_1\lambda_2|}\left(\begin{array}{rr}\xi_1 & \\ & \xi_2\end{array}\right),\ \xi_1=\frac{\lambda_1}{\sqrt{|\lambda_1\lambda_2|}},\ \xi_2=\frac{\lambda_2}{\sqrt{|\lambda_1\lambda_2|}},
\end{equation}
where $\xi_1$ and $\xi_2$ are the normalized eigenvalues of the $\mathcal{R}$-matrix, the corresponding Racah matrix is
\begin{equation}
U=\left(\begin{array}{cc}
-\frac{1}{\xi_1-\xi_2} & \frac{\sqrt{\xi_1^2+1+\xi_2^2}}{\xi_1-\xi_2} \\
\frac{\sqrt{\xi_1^2+1+\xi_2^2}}{\xi_1-\xi_2} & \frac{1}{\xi_1-\xi_2}
\end{array}\right).
\label{1cEH}
\end{equation}

In \cite{AMcabling} a similar conjecture was made for links, i.e. for the three-strand braids with different representations on different strands. Below we discuss the eigenvalue conjecture in two cases: for two-component and three component links. Let components be colored with representations $R_1$, $R_2$ and (for three-component links) $R_3$. In the case of two components, there are two different sets of Racah coefficients: $U_{112}$ and $U_{121}$, which depend on the placement of different components in the braid, as well as there are two diagonal $\mathcal{R}$-matrices: $\mathcal{R}_{11}$ and $\mathcal{R}_{12}$. In the case of links, it is more useful to solve not the Yang-Baxter equations themselves, but the equations which emerge from the cabling procedure (see \cite{AMcabling} for details). In the two component case, the Racah matrices satisfy the equations
\begin{equation}
\begin{array}{l}
U_{121}\mathcal{R}_{12}U_{112}^{\dagger}\mathcal{R}_{11} U_{112}=\mathcal{R}_{(12)\times 1},
\\
U_{112}\mathcal{R}_{12}U_{121}\mathcal{R}_{12} U_{112}^{\dagger}=\mathcal{R}_{(11)\times 1}.
\end{array}
\end{equation}
In contrast with (\ref{YBU}), at the right hand side here there are new $\mathcal{R}$-matrices acting in higher representations, and they appear due to the cabling procedure application. As such, their form is a kind of unknown variable here, although it is known that they should be diagonal and this property is in fact enough to solve these equations. If one expresses the $\mathcal{R}$-matrices through one eigenvalue each\footnote{
The eigenvalues are chosen as being of different signs due to the legacy reasons: for low representations, they usually have different signs. However, the answer actually works for eigenvalues of the same sign as well. One should choose the corresponding normalized eigenvalue as being imaginary, and it can be checked that the formulae still give correct link polynomials (see discussion in the previous section).
 } (this can be done for the normalized $\mathcal{R}$-matrices):
\begin{equation}
\mathcal{R}_{11}=\left(\begin{array}{rr}\xi_{11} & \\ & -\xi^{-1}_{11}\end{array}\right),\ \mathcal{R}_{12}=\left(\begin{array}{rr}\xi_{12} & \\ & -\xi^{-1}_{12}\end{array}\right),
\end{equation}
the Racah matrices are as follows:
\begin{equation}
\begin{array}{l}
U_{112}=\left(\begin{array}{cc}
-\sqrt{\cfrac{(\xi_{11}^2-\xi_{12}^4)}{(1-\xi_{12}^4)(1+\xi_{11}^2)}} & \sqrt{\cfrac{(1-\xi_{11}^2\xi_{12}^4)}{(1-\xi_{12}^4)(1+\xi_{11}^2)}}
\\ \sqrt{\cfrac{(1-\xi_{11}^2\xi_{12}^4)}{(1-\xi_{12}^4)(1+\xi_{11}^2)}} & \sqrt{\cfrac{(\xi_{11}^2-\xi_{12}^4)}{(1-\xi_{12}^4)(1+\xi_{11}^2)}}
\end{array}\right),
\\ \\
U_{121}=\left(\begin{array}{cc}
-\cfrac{\xi_{12}^2(1-\xi_{11}^2)}{\xi_{11}(1-\xi_{12}^4)} & \cfrac{\sqrt{(\xi_{11}^2-\xi_{12}^4)(1-\xi_{12}^4\xi_{11}^2)}}{\xi_{11}(1-\xi_{12}^4)}
\\
\cfrac{\sqrt{(\xi_{11}^2-\xi_{12}^4)(1-\xi_{12}^4\xi_{11}^2)}}{\xi_{11}(1-\xi_{12}^4)} & \cfrac{\xi_{12}^2(1-\xi_{11}^2)}{\xi_{11}(1-\xi_{12}^4)}
\end{array}\right).
\end{array}
\label{2cEH}
\end{equation}

Similarly for three different representations, there are three different Racah matrices: $U_{123}$, $U_{132}$ and $U_{213}$, as  well as three $\mathcal{R}$-matrices: $\mathcal{R}_{12}$, $\mathcal{R}_{13}$ and $\mathcal{R}_{23}$. Three equations defining the Racah coefficients are
\begin{equation}
\begin{array}{l}
U_{123}          \mathcal{R}_{23}U_{132}^{\dagger}\mathcal{R}_{13} U_{213}^{\dagger}=\mathcal{R}_{(12)\times 3},
\\
U_{132}          \mathcal{R}_{23}U_{123}^{\dagger}\mathcal{R}_{12} U_{213}          =\mathcal{R}_{(13)\times 2},
\\
U_{123}^{\dagger}\mathcal{R}_{12}U_{213}          \mathcal{R}_{13} U_{132}          =\mathcal{R}_{(23)\times 1}.
\end{array}
\end{equation}
These equations give the following Racah coefficients:
\begin{equation}
U_{123}=\left(\begin{array}{cc}
-\sqrt{\cfrac{(\xi_{12}^2-\xi_{13}^2\xi_{23}^2)(\xi_{23}^2-\xi_{12}^2\xi_{13}^2)}{\xi_{13}^2(1-\xi_{12}^4)(1-\xi_{23}^4)}} &
\sqrt{\cfrac{(\xi_{13}^2-\xi_{12}^2\xi_{23}^2)(1-\xi_{12}^2\xi_{13}^2\xi_{23}^2)}{\xi_{13}^2(1-\xi_{12}^4)(1-\xi_{23}^4)}}
\\
\sqrt{\cfrac{(\xi_{13}^2-\xi_{12}^2\xi_{23}^2)(1-\xi_{12}^2\xi_{13}^2\xi_{23}^2)}{\xi_{13}^2(1-\xi_{12}^4)(1-\xi_{23}^4)}} &
\sqrt{\cfrac{(\xi_{12}^2-\xi_{13}^2\xi_{23}^2)(\xi_{23}^2-\xi_{12}^2\xi_{13}^2)}{\xi_{13}^2(1-\xi_{12}^4)(1-\xi_{23}^4)}}
\end{array}\right),
\label{3cEH}
\end{equation}
and the remaining two matrices can be easily constructed from this one by simply interchanging the indices. One can definitely easily check that (\ref{1cEH}) is a particular example of (\ref{2cEH}), when the two representations coincide, as well as (\ref{2cEH}) is a particular case of (\ref{3cEH}).

\subsubsection*{$\underline{[1]\otimes[1]\otimes[2]}$}

Since the matrix in section \ref{ex112} is of size $2\times 2$, one can also use the eigenvalue hypothesis in order to calculate it. In this case, the $\mathcal{R}$-matrices in the diagonal form are

\begin{equation}
\mathcal{R}_{[1][1]}=\left(\begin{array}{rr}q & \\ & -q^{-1}\end{array}\right),\ \mathcal{R}_{12}=\sqrt{q}\left(\begin{array}{rr}q^{3/2} & \\ & -q^{-3/2}\end{array}\right),
\end{equation}
then
\begin{equation}
U_{[1][1][2]}=\left(\begin{array}{cc}
-\sqrt{\cfrac{q^2-q^6}{(1-q^6)(1+q^2)}} & \sqrt{\cfrac{(1-q^8)}{(1-q^6)(1+q^2)}}
\\ \sqrt{\cfrac{(1-q^8)}{(1-q^6)(1+q^2)}} & \sqrt{\cfrac{(q^2-q^6)}{(1-q^6)(1+q^2)}}
\end{array}\right)
=\left(\begin{array}{cc}
-\sqrt{\cfrac{1}{[3]}} & \sqrt{\cfrac{[4]}{[2][3]}}
\\ \sqrt{\cfrac{[4]}{[2][3]}} & \sqrt{\cfrac{1}{[3]}}
\end{array}\right).
\end{equation}
This answer definitely coincides with the one calculated using representation theory in (\ref{u112ans}).

\section{HOMFLY polynomials for multicolored links\label{s.HOMFLY}}

Using these calculated inclusive multicolored Racah matrices, one can evaluate the multicolored HOMFLY polynomials for links that have three-strand braid presentations. Here we discuss some simple examples: the Whitehead link, the Borromean rings and link L7a3 in the Thistlethwaite Link Table \cite{twi}.

\subsection{Whitehead link}

The Whitehead link is a two-component link. It has the following braid representation:

\begin{picture}(160,80)(-70,-10)
\put(0,48){\line(1,0){20}}
\put(0,24){\line(1,0){20}}
\put(0,0){\line(1,0){44}}
\put(0,51){$R_2$}
\put(0,27){$R_1$}
\put(0,3){$R_1$}
\put(20,48){\line(1,-1){10}}
\put(44,24){\line(-1,1){10}}
\put(20,24){\line(1,1){24}}
\put(24,46){$\mathcal{R}_{12}$}
\put(44,24){\line(1,-1){24}}
\put(44,0){\line(1,1){10}}
\put(68,24){\line(-1,-1){10}}
\put(44,48){\line(1,0){24}}
\put(48,-3){$\mathcal{R}_{12}$}
\put(68,48){\line(1,-1){10}}
\put(92,24){\line(-1,1){10}}
\put(68,24){\line(1,1){24}}
\put(68,0){\line(1,0){24}}
\put(72,46){$\mathcal{R}_{11}$}
\put(92,24){\line(1,-1){24}}
\put(92,0){\line(1,1){10}}
\put(116,24){\line(-1,-1){10}}
\put(92,48){\line(1,0){24}}
\put(96,-3){$\mathcal{R}_{12}$}
\put(116,48){\line(1,-1){10}}
\put(140,24){\line(-1,1){10}}
\put(116,24){\line(1,1){24}}
\put(118,46){$\mathcal{R}_{12}$}
\put(116,0){\line(1,0){44}}
\put(140,24){\line(1,0){20}}
\put(140,48){\line(1,0){20}}
\end{picture}

Since it has two components, one needs two $\mathcal{R}$- and Racah matrices. As we explained above, there are two diagonal $\mathcal{R}$-matrices: $\mathcal{R}_{12}$ that stands for crossings of representations $R_1$ and $R_2$,  and $\mathcal{R}_{11}$ that stands for crossing between representations $R_1$ and $R_1$. The two Racah matrices $U_{112}$ and $U_{121}$ correspond accordingly to the placements of representations $R_1R_1R_2$  and  $R_1R_2R_1$.

The answer for the HOMFLY polynomial is then given by the character expansion \cite{MMMI}:
\begin{equation}
H_{R_1R_2}^{\text{Whitehead}}=\sum\limits_{Q\vdash R_1\otimes R_1\otimes R_2} S_Q^*(A,q) B_Q,
\end{equation}
where $S_Q^*(A,q)$ is the quantum dimension\footnote{
For these quantities, there is a simple hook formula which allows one to easily calculate them:
\begin{equation}
S^*_Q(A,q)=\prod_{(i,j)\in Q}
\frac{Aq^{i-j}-A^{-1}q^{j-i}}{q^{h_{i,j}}-q^{-h_{i,j}}}.
\begin{picture}(105,15)(-35,-15)
\put(0,0){\line(1,0){70}}
\put(0,-10){\line(1,0){70}}
\put(0,-20){\line(1,0){60}}
\put(0,-30){\line(1,0){40}}
\put(0,-40){\line(1,0){20}}
\put(0,-50){\line(1,0){20}}
\put(0,0){\line(0,-1){50}}
\put(10,0){\line(0,-1){50}}
\put(20,0){\line(0,-1){50}}
\put(30,0){\line(0,-1){30}}
\put(40,0){\line(0,-1){30}}
\put(50,0){\line(0,-1){20}}
\put(60,0){\line(0,-1){20}}
\put(70,0){\line(0,-1){10}}
\put(15,-15){\makebox(0,0)[cc]{\textbf{x}}}
\put(15,5){\makebox(0,0)[cc]{$i$}}
\put(-5,-15){\makebox(0,0)[cc]{$j$}}
\qbezier(19,-11)(45,20)(55,-15)
\put(40,10){\makebox(0,0)[cc]{$k$}}
\qbezier(11,-19)(-17,-40)(15,-45)
\put(60,-40){\makebox(0,0)[lc]{$h_{i,j}=k+l+1$}}
\end{picture}
\end{equation}

\noindent $[n]_q$ denotes the quantum number, i.e. $[n]_q\equiv\frac{q^n-q^{-n}}{q-q^{-1}}$.
}
of the representation $Q$ of $SU_q(N)$, and $B_Q$ is a trace of the product of $\mathcal{R}$ and Racah matrices for the representation $Q$:
\begin{equation}
B=Tr\left(\mathcal{R}_{12} U_{121} \mathcal{R}^{-1}_{12} U_{112}^{\dagger} \mathcal{R}_{11} U_{112} \mathcal{R}^{-1}_{12} U_{121} \mathcal{R}_{12} \right).
\end{equation}

The eigenvalues of the diagonal $\mathcal{R}$-matrix depend on the representations appearing in the decomposition of the product
$R_1\otimes R_1$ for $\mathcal{R}_{11}$ and $R_1\otimes R_2$ for $\mathcal{R}_{12}$. For $\mathcal{R}_{12}$, these eigenvalues are generally equal to
\begin{equation}
\lambda_i=q^{\varkappa_{Q_i}-\varkappa_{R_1}-\varkappa_{R_2}}\ \ \text{for } Q_i\vdash R_1\otimes R_2,Q\vdash Q_i\otimes R_1.
\end{equation}
Unfortunately, signs of the eigenvalues cannot be determined uniquely.

First, let us consider the intrinsic causes and repeat in this concrete case the argument of section 5. Since
\begin{equation}
\mathcal{R}_{12}: R_1 \otimes R_2\ \rightarrow\ \  R_2  \otimes R_1,
\end{equation}
$\mathcal{R}_{12}$ is diagonal, if one chooses some $u$ and $v$ as bases in $R_1\otimes R_2$ and $R_2\otimes R_1$ in such a way that
\begin{equation}
\mathcal{R}_{12} \cdot u= \lambda v.
\end{equation}
As clear from this equation, signs of the eigenvalues depend on the choice of signs of $u$ and $v$, and they remain unchanged only when the two bases change their signs together. In this case, by the definition of the rotation matrix, the Racah matrix $U_{121}$ is the same. Therefore, each choice of signs of the eigenvalues corresponds to only one Racah matrix.

If the representation $Q$ has multiplicity 1 in $R_1 \otimes R_2 \otimes R_1$, there is no problem because the Racah matrix is trivial. But for those representations $Q$ of multiplicity $n\geq 2$, things become more complicated.

Let us denote those eigenvalues to be $\lambda_i(1\leq i \leq n)$. If we fix the sign of $\lambda_1$, there are $2^{n-1}$ ways to combine the signs of different eigenvalues, and it is possible to have several combinations resulting in the same Racah matrix. However, in the highest weight vector method, the Racah matrix is obtained uniquely without any references to $\mathcal{R}$ matrix, so the complexity lies in the determination of the ``right'' $\mathcal{R}$ matrices among those possibilities with the unique Racah matrix.

Actually, as we explained earlier, one good thing is that the signs of eigenvalues of knots are easier to be determined. If one studies knots in representation $R=R_1=R_2$, each irreducible representation $Q_i\vdash R\otimes R$ lies either in the symmetric or antisymmetric square, which determines the sign of $\lambda_Q$: it is plus if $Q$ belongs to the symmetric square and minus otherwise. See \cite{MMMS21} for more details.

These formulae together with the Racah matrices calculated in the present paper are enough to evaluate several (multi)colored HOMFLY polynomials for the Whitehead link. Most of them are given in Appendix \ref{A.w}. Here we present only one non-trivial example:

\begin{equation}
\begin{array}{r}
\{A\}H_{[2,1][1]}=A^{-2}\big(q^{4}-q^{2}-q^{-2}+q^{-4}\big)+\big(-q^{8}+2q^{6}-3q^{4}+4q^{2}-5+4q^{-2}-3q^{-4}+2q^{-6}-q^{-8}\big)+
\\ \\
A^{2}\big(q^{4}-q^{2}+1-q^{-2}+q^{-4}\big).
\end{array}
\end{equation}

\subsubsection*{Properties of the answers}

These polynomials possess several properties which can be used to check the answer.

The main property of the link polynomials is that if one ``normalizes'' the answer by dividing it by the quantum dimensions of the representations on different components of the link then this normalized answer has no quantum numbers in the denominator, only terms of the form $(Aq^i-A^{-1}q^{-i})$ remain there. For the Whitehead link, the ``normalized'' polynomial is defined as
\begin{equation}
\mathcal{H}_{Q_1,Q_2}=\frac{H_{Q_1,Q_2}}{S_{Q_1}^*S_{Q_2}^*}.
\end{equation}

The second property of the Whitehead invariant is that if one takes a particular group $SU(N)$ (i.e. puts $A=q^N$) and if representation $Q_1$ is trivial in this group (as it happens for representation $[1,1]$ in $SU(2)$ and for representation $[1,1,1]$ in $SU(3)$), then the answer for the HOMFLY polynomial becomes just the answer for the unknot in representation $Q_2$:
\begin{equation}
H_{[r^N],Q_2}|_{A=q^N}=S^*_{Q_2}(A=q^N,q).
\end{equation}
The same is of course true if $Q_2$ is trivial, then the result is the quantum dimension of representation $Q_1$. This property is only true for the links with all components unknotted by themselves. Otherwise, the result would be the HOMFLY polynomial of the component rather than the polynomial of the unknot (see a detailed discussion of these properties in \cite{BJLMMMS}).

The third property, which is general for all knots and links is that the transposition of Young diagrams corresponds to the substitution $q\rightarrow -q^{-1}$. For example:
\begin{equation}
\begin{array}{l}
H_{[2],[2]}(A,q)=H_{[1,1],[1,1]}(A,-q^{-1});
\\
H_{[3],[2,1]}(A,q)=H_{[1,1,1],[2,1]}(A,-q^{-1});
\\
\text{etc.}
\end{array}
\end{equation}

The fourth property is peculiar for the Whitehead link. Although it is not obvious from the braid picture, this link is actually symmetric under permutation of its components meaning that $H_{R_1R_2}=H_{R_2R_1}$.

\subsection{Borromean rings}

The second example is the Borromean rings. The braid representation of this link is

\begin{picture}(160,80)(-70,-10)
\put(0,48){\line(1,0){20}}
\put(0,24){\line(1,0){20}}
\put(0,0){\line(1,0){44}}
\put(0,51){$R_1$}
\put(0,27){$R_2$}
\put(0,3){$R_3$}
\put(20,48){\line(1,-1){10}}
\put(44,24){\line(-1,1){10}}
\put(20,24){\line(1,1){24}}
\put(24,46){$\mathcal{R}_{12}$}
\put(44,24){\line(1,-1){24}}
\put(44,0){\line(1,1){10}}
\put(68,24){\line(-1,-1){10}}
\put(44,48){\line(1,0){24}}
\put(48,-3){$\mathcal{R}_{13}$}
\put(68,48){\line(1,-1){10}}
\put(92,24){\line(-1,1){10}}
\put(68,24){\line(1,1){24}}
\put(68,0){\line(1,0){24}}
\put(72,46){$\mathcal{R}_{23}$}
\put(92,24){\line(1,-1){24}}
\put(92,0){\line(1,1){10}}
\put(116,24){\line(-1,-1){10}}
\put(92,48){\line(1,0){24}}
\put(96,-3){$\mathcal{R}_{12}$}
\put(116,48){\line(1,-1){10}}
\put(140,24){\line(-1,1){10}}
\put(116,24){\line(1,1){24}}
\put(116,0){\line(1,0){24}}
\put(118,46){$\mathcal{R}_{13}$}
\put(140,24){\line(1,-1){24}}
\put(140,0){\line(1,1){10}}
\put(164,24){\line(-1,-1){10}}
\put(144,-3){$\mathcal{R}_{23}$}
\put(164,0){\line(1,0){20}}
\put(164,24){\line(1,0){20}}
\put(140,48){\line(1,0){44}}
\end{picture}

It has three components, thus, one needs three $\mathcal{R}$- and Racah matrices. There are three diagonal $\mathcal{R}$-matrices: $\mathcal{R}_{12}$, $\mathcal{R}_{13}$ and $\mathcal{R}_{23}$ (see also section \ref{incrac}). Similarly, there are three Racah matrices\footnote{$U_{123}$ here corresponds to the Racah matrix which transforms from the basis $(R_1\otimes R_2)\otimes R_3$ to the basis $R_1\otimes (R_2\otimes R_3)$. Thus $U_{321}=U_{123}^{\dagger}$.}: $U_{123}$, $U_{132}$ and $U_{213}$.

The answer for the HOMFLY polynomial is then given by formula \cite{MMMI}:
\begin{equation}
H_{R_1R_2R_3}^{\text{Borromean}}=\sum\limits_{Q\vdash R_1\otimes R_2\otimes R_3} S_Q^*(A,q) B_Q,
\end{equation}
$B_Q$ here is given by the trace of the product of $\mathcal{R}$- and Racah matrices in representation $Q$:
\begin{equation}
B=Tr\left(\mathcal{R}_{12} U_{213} \mathcal{R}^{-1}_{13} U_{132} \mathcal{R}_{23} U_{123}^{\dagger} \mathcal{R}^{-1}_{12} U_{213}^{\dagger} \mathcal{R}_{13} U_{132} \mathcal{R}^{-1}_{23} U_{123}^{\dagger} \right).
\end{equation}

The eigenvalues of the diagonal $\mathcal{R}$-matrix depend on the representations appearing in the decomposition of the product $R_1\otimes R_1$ for $\mathcal{R}_{11}$ and $R_1\otimes R_2$ for $\mathcal{R}_{12}$. Generally these eigenvalues are equal to
\begin{equation}
\lambda_i=q^{\varkappa_{Q_i}-\varkappa_{R_1}-\varkappa_{R_2}}\ \ \text{for}\ Q_i\vdash R_1\otimes R_2.
\end{equation}
For the same reason as in the case of the Whitehead link, their signs are not uniquely defined.

These formulae together with the Racah matrices calculated in this paper are enough to evaluate several (multi)colored HOMFLY polynomials for the Borromean rings. Most of them are given in Appendix \ref{A.b}. Here we present only one non-trivial example:
\begin{equation}
\begin{array}{l}
\{A\}^2H_{[2,1],[1],[1]}=\big(q^{10}-4q^{8}+8q^{6}-12q^{4}+15q^{2}-18+15q^{-2}-12q^{-4}+8q^{-6}-4q^{-8}+q^{-10}\big)+
\\ \\
+A^{-2}\big(-q^{6}+3q^{4}-3q^{2}+3-3q^{-2}+3q^{-4}-q^{-6}\big)+A^{2}\big(-q^{6}+3q^{4}-3q^{2}+3-3q^{-2}+3q^{-4}-q^{-6}\big).
\end{array}
\end{equation}

\subsubsection*{Properties of the answers}

These polynomials possess several properties which can be used to check the answer.

For Borromean rings, the ``normalized'' polynomial is defined as
\begin{equation}
\mathcal{H}_{Q_1,Q_2,Q_3}=\frac{H_{Q_1,Q_2,Q_3}}{S_{Q_1}^*S_{Q_2}^*S_{Q_3}^*},
\end{equation}
and its main property is again that it has no quantum numbers in the denominator, only terms of the form $(Aq^i-A^{-1}q^{-i})$ remain there.

The second property is again behaviour under the transposition of Young diagrams. For example:
\begin{equation}
\begin{array}{l}
H_{[2],[2],[1,1]}(A,q)=H_{[1,1],[1,1],[2]}(A,-q^{-1});
\\
H_{[3],[2,1],[1,1,1]}(A,q)=H_{[1,1,1],[2,1],[3]}(A,-q^{-1});
\\
\text{etc.}
\end{array}
\end{equation}

The third property is peculiar for the Borromean rings: it is symmetric under permutating any of its components, e.g. $H_{R_1R_2R_3}=H_{R_2R_1R_3}$ etc.

\subsection{L7a3}
The third example is link L7a3 in the Thistlethwaite link table \cite{twi}. It is a two-component link, which has a three-strand representation

\begin{picture}(160,80)(-70,-10)
\put(0,48){\line(1,0){18}}
\put(0,24){\line(1,0){18}}
\put(0,0){\line(1,0){34}}
\put(0,51){$R_1$}
\put(0,27){$R_2$}
\put(0,3){$R_2$}
\put(18,48){\line(1,-1){30}}
\put(18,24){\line(1,1){10}}
\put(34,0){\line(1,1){24}}
\put(22,46){$\mathcal{R}_{12}$}
\put(42,48){\line(-1,-1){10}}
\put(64,0){\line(-1,1){12}}
\put(64,24){\line(1,0){8}}
\put(42,2){$\mathcal{R}_{12}$}
\put(42,48){\line(1,0){30}}
\put(64,0){\line(1,0){52}}
\put(57,24){\line(1,0){16}}
\put(72,48){\line(1,-1){24}}
\put(72,24){\line(1,1){10}}
\put(96,48){\line(-1,-1){10}}
\put(76,46){$\mathcal{R}_{22}$}
\put(42,48){\line(1,0){30}}
\put(96,48){\line(1,0){10}}
\put(96,24){\line(1,0){10}}
\put(106,24){\line(1,1){10}}
\put(106,48){\line(1,-1){24}}
\put(110,46){$\mathcal{R}_{22}$}
\put(130,48){\line(-1,-1){10}}
\put(130,48){\line(1,0){10}}
\put(130,24){\line(1,0){10}}
\put(143,46){$\mathcal{R}_{22}$}
\put(140,48){\line(1,-1){28}}
\put(140,24){\line(1,1){10}}
\put(164,48){\line(-1,-1){10}}
\put(164,48){\line(1,0){10}}
\put(174,48){\line(1,-1){24}}
\put(198,24){\line(1,0){30}}
\put(64,0){\line(1,0){90}}
\put(154,0){\line(1,1){32}}
\put(188,0){\line(-1,1){15}}
\put(188,0){\line(1,0){41}}
\put(201,48){\line(-1,-1){12}}
\put(201,48){\line(1,0){28}}
\put(178,46){$\mathcal{R}_{12}$}
\put(163,2){$\mathcal{R}_{12}$}
\end{picture}

It has two components, thus one needs two $\mathcal{R}$- and Racah matrices. There are two diagonal $\mathcal{R}$-matrices: $\mathcal{R}_{12}$ that stands for crossings of representations $R_1$ and $R_2$  and $\mathcal{R}_{22}$ that stands for crossings of representations $R_2$ and $R_2$  (see also section \ref{incrac}). The two Racah matrices $U_{122}$ and $U_{212}$ correspond accordingly to the placements of representations $R_1R_2R_2$ and $R_2R_1R_2$ .

The answer for the HOMFLY polynomial is then given by the character expansion \cite{MMMI}:
\begin{equation}
H_{R_1R_2}^{\text{L7a3}}=\sum\limits_{Q\vdash R_1\otimes R_2\otimes R_2} S_Q^*(A,q) B_Q,
\end{equation}
$B_Q$ here is given by the trace of the following product of $\mathcal{R}$ and Racah matrices for representation $Q$:
\begin{equation}
B=Tr\left(\mathcal{R}_{12} U_{212} \mathcal{R}^{-1}_{12} U_{122} \mathcal{R}^{3}_{22} U_{122}^{\dagger} \mathcal{R}^{-1}_{12} U_{212} \mathcal{R}_{12} \right).
\end{equation}

The eigenvalues of the diagonal $\mathcal{R}$-matrix depend on the representations appearing in the decomposition of the product
$R_1\otimes R_2$ for $\mathcal{R}_{12}$ and $R_2\otimes R_2$ for $\mathcal{R}_{22}$. Generally this eigenvalues are equal to
\begin{equation}
\lambda_i=q^{\varkappa_{Q_i}-\varkappa_{R_1}-\varkappa_{R_2}}\ \ \text{for}\ Q_i\vdash R_1\otimes R_2.
\end{equation}
These formulae together with the Racah matrices calculated in this paper are enough to calculate several (multi)colored HOMFLY polynomials for link L7a3. Most of them are given in Appendix \ref{A.l}. Here we present only one non-trivial example:
\begin{equation}
\begin{array}{r}
\{A\}H_{[3],[1]}=A^{-2}\big(q^{4}-q^{2}+2-2q^{-2}+q^{-4}-q^{-6}+q^{-8}\big)+A^{2}\big(q^{8}-q^{6}+q^{4}-q^{2}+1+q^{-4}\big)+
\\ \\
+\big(-q^{10}+q^{8}-q^{6}+2q^{4}-3q^{2}+1-3q^{-2}+2q^{-4}-q^{-6}+q^{-8}-q^{-10}\big).
\end{array}
\end{equation}

\subsubsection*{Properties of the answers}

These polynomials possess several properties which can be used to check the answer.

For link L7a3, the ``normalized'' polynomial is defined as
\begin{equation}
\mathcal{H}_{Q_1,Q_2}=\frac{H_{Q_1,Q_2}}{S_{Q_1}^*S_{Q_2}^*},
\end{equation}
and its main property is again that it has no quantum numbers in the denominator, only terms of the form $(Aq^i-A^{-1}q^{-i})$ remain there.

The second property of the L7a3 polynomial is that if one takes a particular group $SU(N)$ (i.e. puts $A=q^N$) and if representation $Q_1$ is trivial in this group (as it happens for representation $[1,1]$ in the $SU(2)$ and $[1,1,1]$ in representation $SU(3)$), then the answer for the HOMFLY polynomial becomes just the answer for the unknot in representation $Q_2$:
\begin{equation}
H_{[r^N],Q_2}|_{A=q^N}=S^*_{Q_2}(A=q^n,q).
\end{equation}
The same is of course true if $Q_2$ is trivial, then the result is the quantum dimension of representation $Q_1$. This property is only true for the links with all components unknotted by themselves. Otherwise, the result would be HOMFLY polynomial of the component rather than the polynomial of the unknot.

The third property is again the behaviour under the transposition of Young diagrams. For example:
\begin{equation}
\begin{array}{l}
H_{[2],[2]}(A,q)=H_{[1,1],[1,1]}(A,-q^{-1});
\\
H_{[2,1],[3]}(A,q)=H_{[2,1],[1,1,1]}(A,-q^{-1});
\\
\text{etc.}
\end{array}
\end{equation}

\section{Exclusive Racah matrices\label{s.exex}}

Since this paper is devoted mainly to evaluating the inclusive Racah matrices, we will not delve deep into the story of the exclusive ones. However, we will present one example of the calculation of exclusive link Racah matrix using the trick described in s.\ref{exrac}, namely we will discuss the matrix $S_{\bar{[1]},[1],[2]}$.

To use the trick, one should first calculate the HOMFLY polynomial for the two-parametric series of links
\begin{equation*}
H_{[2],[1],[1]}^{(2m,-1|2n,\pm 1)}\!\!\!\!\!=\!\!\!\!\!\!\!\!\!\!\sum\limits_{Q\vdash [2]\otimes [1]\otimes [1]}\!\!\!\!\!\!\!\!\!\! S^*_Q \text{Trace}\left(
\mathcal{R}^{2m}_{[1],[2]} U_{[1],[1],[2]}^{\dagger} \mathcal{R}^{-1}_{[1],[1]} U_{[1],[1],[2]} \mathcal{R}^{2n}_{[1],[2]} U_{[1],[1],[2]}^{\dagger} \mathcal{R}^{-1}_{[1],[1]} U_{[1],[1],[2]}
\right)_Q\!\!\!=\!\!\!\!\!\sum\limits_{{Y\vdash [1]\otimes [2]}\atop{Z\vdash [1]\otimes [2]}}\!\!\!\!\! h_{YZ}\lambda_Y^{2m}\lambda_Z^{2n}.
\end{equation*}
The Racah matrices needed for this calculation one can get also from the eigenvalue hypothesis, see s.\ref{s.eig}. This leads to a matrix of coefficients $h_{YZ}$:
\begin{equation}
h_{YZ}=\begin{array}{c|cc}
Y\backslash Z & [3] & [2,1]
\\
\hline
\\
\ [3] & \cfrac{\{Aq^2\}(A^3q^2-Aq^4+Aq^2-Aq^{-4})}{\{A\}^2[3]_q^2} & \cfrac{\{Aq^2\}\{Aq^{-1}\}[2]_q}{\{A\}^2[3]_q^2}
\\
\ [2,1] & \cfrac{\{Aq^2\}\{Aq^{-1}\}[2]_q}{\{A\}^2[3]_q^2} & \cfrac{\{Aq^{-1}\}[2]_q(A^3+A^3q^{-2}-Aq^4-A+Aq^{-2}-Aq^{-4})}{\{A\}^2[3]_q^2}
\end{array}
\end{equation}
After dividing its elements by the square roots of dimensions $h_{YZ}/\sqrt{d_Y d_Z}$, the resulting matrix can be diagonalized by the matrix $S_{\bar{[1]},[1],[2]}$:
\begin{equation}
S_{\bar{[1]},[1],[2];\bar{X}_1Y}\cfrac{h_{YZ}}{\sqrt{d_Y d_Z}}S_{\bar{[1]},[1],[2];\bar{X}_2Z}^{\dagger}=\text{diagonal matrix}_{\bar{X}_1,\bar{X}_2}.
\end{equation}
Since the Racah matrix is unitary, this is a set of linear equations for $S_{\bar{[1]},[1],[2]}$, the solution being
\begin{equation}
S_{\bar{[1]},[1],[2];\bar{X}Y}=\begin{array}{c|cc}
\bar{X}\backslash Y & [3] & [2,1] \\\hline\\
\varnothing & -\sqrt{\cfrac{\{Aq\}\{A\}\{Aq^{-1}\}}{[3]_q}} & \sqrt{\cfrac{\{Aq^2\}\{Aq\}\{A\}}{[2]_q[3]_q}}
\\
\text{adjoint} & \sqrt{\cfrac{\{Aq^2\}\{Aq\}\{A\}}{[2]_q[3]_q}} & \sqrt{\cfrac{\{Aq\}\{A\}\{Aq^{-1}\}}{[3]_q}}
\end{array}
\end{equation}
This answer is in perfect agreement with the answer obtained in \cite{RacahRama}.

\section{Conclusion}
 In this paper, we have calculated the inclusive Racah matrices for all representations up to size $3$ using the highest weight method. Using these matrices, we have evaluated the HOMFLY polynomials for the Whitehead link, Borromean rings and link L7a3. Also calculated inclusive Racah matrices allowed us to check the earlier suggested method of calculation of the exclusive Racah matrices from the inclusive ones: we have evaluated the exclusive Racah matrix $S_{\bar{[1]},[1],[2]}$ and checked that it coincides with the known answer for this matrix.

\section*{Acknowledgements}
The work was partly supported by the grant of the Foundation for the Advancement of Theoretical Physics “BASIS” (A.Mor. and A.S.), by RFBR grants 16-01-00291 (A.Mir.), 16-02-01021 (A.Mor.), 17-01-00585 (An.Mor.) and 16-31-60082-mol-a-dk (A.S.), by joint grants 17-51-50051-YaF, 15-51-52031-NSC-a, 16-51-53034-GFEN, 16-51-45029-IND-a (A.M.’s and A.S.), by grant NSFC (11425104,11611130015) (C.B., J.J. and J.L.) the grant of Ningde Normal University (Grant No. 2017Y07) (J.J.). Chengming Bai, Jianjian Jiang and Jinting Liang thank ITEP for the hospitality and for useful discussions while they visited there in the summer of 2017.

\appendix

\section{Link Polynomials\label{a.pol}}

In this Appendix, we list the answers for the normalized link polynomials calculated using the Racah coefficients provided in this paper. The normalized polynomials are made from unnormalized ones by dividing them by dimensions:
\begin{equation}
\mathcal{H}_{1,2}=\cfrac{H_{1,2}}{d_1 d_2},\ \ \ \ \ \mathcal{H}_{1,2,3}=\cfrac{H_{1,2,3}}{d_1 d_2 d_3}.
\end{equation}
We use the notation for them in the matrix form suggested in \cite{Inds8}. The matrix describes
the coefficients of a polynomial in $A^2$ and $q^2$ as in the following example:

\setlength{\arraycolsep}{1pt}

\begin{equation}
\nonumber
 q^{10}A^{16}+2q^{12}A^{16}+3q^{10}A^{18}+4q^{12}A^{18}\longrightarrow
q^{10}A^{16} \left(\begin{array}{rr}
3 & 4 \\
& \\
1 & 2 \\
\end{array}\right).
\end{equation}

\subsection{Whitehead link\label{A.w}}

\begin{equation}
\mathcal{H}_{[1],[1]}=\frac{1}{q^{4}A^{2}\{A\}}
\left(\begin{array}{rrrrr}
0 & 1 & -1 & 1 & 0 \\
&&&& \\
-1 & 2 & -3 & 2 & -1 \\
&&&& \\
0 & 1 & -2 & 1 & 0
\end{array}\right).
\end{equation}
\begin{equation}
\mathcal{H}_{[1],[2]}=\frac{1}{q^{6}A^{2}\{A\}}
\left(\begin{array}{rrrrrrr}
0 & 0 & 1 & 0 & -1 & 1 & 0 \\
&&&&&& \\
-1 & 1 & 1 & -3 & 1 & 1 & -1 \\
&&&&&& \\
0 & 1 & -1 & -1 & 1 & 0 & 0
\end{array}\right).
\end{equation}
\begin{equation}
\mathcal{H}_{[2],[2]}=\frac{1}{q^{11}A^{4}\{A\}\{Aq\}}
\left(\begin{array}{rrrrrrrrrrrrr}
0 & 0 & 0 & 0 & 0 & 1 & -1 & 0 & 2 & -1 & -1 & 1 & 0 \\
&&&&&&&&&&&& \\
0 & 0 & -1 & 1 & 1 & -4 & 1 & 4 & -4 & -2 & 3 & 0 & -1 \\
&&&&&&&&&&&& \\
1 & -2 & 1 & 4 & -7 & 0 & 10 & -6 & -5 & 6 & 0 & -2 & 1 \\
&&&&&&&&&&&& \\
-1 & 1 & 3 & -5 & -2 & 8 & -2 & -5 & 3 & 1 & -1 & 0 & 0 \\
&&&&&&&&&&&& \\
0 & 1 & -2 & -1 & 4 & -1 & -2 & 1 & 0 & 0 & 0 & 0 & 0
\end{array}\right).
\end{equation}
\begin{equation}
\mathcal{H}_{[2],[1,1]}=\frac{1}{q^{6}A^{1}\{A\}}
\left(\begin{array}{rrrrrrrrr}
0 & 0 & 1 & 0 & -1 & 0 & 1 & 0 & 0 \\
&&&&&&&& \\
-1 & 0 & 2 & 0 & -3 & 0 & 2 & 0 & -1 \\
&&&&&&&& \\
0 & 0 & 1 & 0 & -2 & 0 & 1 & 0 & 0
\end{array}\right).
\end{equation}
\begin{equation}
\mathcal{H}_{[1],[3]}=\frac{1}{q^{8}A^{2}\{A\}}
\left(\begin{array}{rrrrrrrrr}
0 & 0 & 0 & 1 & 0 & 0 & -1 & 1 & 0 \\
&&&&&&&& \\
-1 & 1 & 0 & 1 & -3 & 1 & 0 & 1 & -1 \\
&&&&&&&& \\
0 & 1 & -1 & 0 & -1 & 1 & 0 & 0 & 0
\end{array}\right).
\end{equation}
\begin{equation}
\mathcal{H}_{[1],[2,1]}=\frac{1}{q^{8}A^{2}\{A\}}
\left(\begin{array}{rrrrrrrrr}
0 & 0 & 1 & -1 & 1 & -1 & 1 & 0 & 0 \\
&&&&&&&& \\
-1 & 2 & -3 & 4 & -5 & 4 & -3 & 2 & -1 \\
&&&&&&&& \\
0 & 0 & 1 & -1 & 0 & -1 & 1 & 0 & 0
\end{array}\right).
\end{equation}
\begin{equation}
\mathcal{H}_{[2],[3]}=\frac{1}{q^{15}A^{4}\{A\}\{Aq\}}
\left(\begin{array}{rrrrrrrrrrrrrrrrr}
0 & 0 & 0 & 0 & 0 & 0 & 0 & 1 & 0 & -1 & 0 & 1 & 1 & -1 & -1 & 1 & 0 \\
&&&&&&&&&&&&&&&& \\
0 & 0 & 0 & -1 & 0 & 2 & 0 & -3 & -2 & 3 & 3 & -3 & -3 & 1 & 2 & 0 & -1 \\
&&&&&&&&&&&&&&&& \\
1 & -1 & -2 & 3 & 3 & -3 & -6 & 2 & 9 & -1 & -7 & -1 & 4 & 2 & -2 & -1 & 1 \\
&&&&&&&&&&&&&&&& \\
-1 & 0 & 3 & 1 & -5 & -3 & 5 & 5 & -3 & -5 & 1 & 3 & 0 & -1 & 0 & 0 & 0 \\
&&&&&&&&&&&&&&&& \\
0 & 1 & -1 & -2 & 1 & 2 & 1 & -2 & -1 & 1 & 0 & 0 & 0 & 0 & 0 & 0 & 0
\end{array}\right).
\end{equation}
\begin{equation}
\mathcal{H}_{[2],[2,1]}=\frac{1}{q^{13}A^{4}\{A\}\{Aq\}}
\left(\begin{array}{rrrrrrrrrrrrrrr}
0 & 0 & 0 & 0 & 0 & 1 & -1 & 0 & 1 & 0 & 0 & -1 & 1 & 0 & 0 \\
&&&&&&&&&&&&&& \\
0 & 0 & -1 & 1 & 0 & -1 & 0 & -1 & 2 & -1 & -1 & 0 & 0 & 1 & -1 \\
&&&&&&&&&&&&&& \\
1 & -2 & 2 & -1 & 1 & -1 & -1 & 5 & -4 & 0 & -1 & 2 & 1 & -2 & 1 \\
&&&&&&&&&&&&&& \\
-1 & 2 & -1 & 1 & -3 & 0 & 4 & -1 & 0 & -3 & 2 & 1 & -1 & 0 & 0 \\
&&&&&&&&&&&&&& \\
0 & 0 & 1 & -2 & 0 & 1 & 1 & 0 & -2 & 1 & 0 & 0 & 0 & 0 & 0
\end{array}\right).
\end{equation}
\begin{equation}
H_{[2],[1,1,1]}=\frac{1}{q^{8}A^{1}\{A\}}
\left(\begin{array}{rrrrrrrrrrr}
0 & 0 & 1 & 0 & -1 & 0 & 0 & 1 & 0 & 0 & 0 \\
&&&&&&&&&& \\
-1 & 0 & 1 & 1 & 0 & -3 & 0 & 1 & 1 & 0 & -1 \\
&&&&&&&&&& \\
0 & 0 & 0 & 1 & 0 & -1 & -1 & 0 & 1 & 0 & 0
\end{array}\right).
\end{equation}
\begin{equation}
\mathcal{H}_{[3],[2,1]}=\frac{1}{q^{13}A^{3}\{A\}\{Aq\}}
\left(\begin{array}{rrrrrrrrrrrrrrrrrrr}
0 & 0 & 0 & 0 & 0 & 0 & 0 & 1 & 0 & -1 & 0 & 0 & 2 & -1 & 0 & -1 & 1 & 0 & 0 \\
&&&&&&&&&&&&&&&&&& \\
0 & 0 & 0 & -1 & 0 & 1 & 1 & -1 & -3 & 1 & 0 & 4 & -4 & 1 & -3 & 3 & -1 & 1 & -1 \\
&&&&&&&&&&&&&&&&&& \\
1 & -1 & -1 & 1 & 1 & 3 & -6 & 1 & -2 & 8 & -2 & -2 & -4 & 2 & 3 & 0 & -1 & -1 & 1 \\
&&&&&&&&&&&&&&&&&& \\
-1 & 1 & 0 & 2 & -1 & -3 & -1 & 1 & 6 & -1 & -3 & -3 & 2 & 2 & 0 & -1 & 0 & 0 & 0 \\
&&&&&&&&&&&&&&&&&& \\
0 & 0 & 1 & -1 & -1 & -1 & 2 & 2 & -1 & -1 & -1 & 1 & 0 & 0 & 0 & 0 & 0 & 0 & 0
\end{array}\right).
\end{equation}
\begin{equation}
\mathcal{H}_{[3],[1,1,1]}=\frac{1}{q^{6}\{A\}}
\left(\begin{array}{rrrrrrrrrrrrr}
0 & 0 & 0 & 1 & 0 & 0 & -1 & 0 & 0 & 1 & 0 & 0 & 0 \\
&&&&&&&&&&&& \\
-1 & 0 & 0 & 2 & 0 & 0 & -3 & 0 & 0 & 2 & 0 & 0 & -1 \\
&&&&&&&&&&&& \\
0 & 0 & 0 & 1 & 0 & 0 & -2 & 0 & 0 & 1 & 0 & 0 & 0
\end{array}\right).
\end{equation}

\newpage
\begin{landscape}
\begin{equation}
\mathcal{H}_{[3],[3]}=\frac{1}{q^{23}A^{6}\{A\}\{Aq\}\{Aq\}}
\left(\begin{array}{rrrrrrrrrrrrrrrrrrrrrrrrrrr}
0 & 0 & 0 & 0 & 0 & 0 & 0 & 0 & 0 & 0 & 0 & 0 & 0 & 1 & -1 & 0 & 1 & 1 & -1 & -2 & 1 & 2 & 0 & -1 & -1 & 1 & 0 \\
&&&&&&&&&&&&&&&&&&&&&&&&&& \\
0 & 0 & 0 & 0 & 0 & 0 & 0 & 0 & -1 & 1 & 0 & 0 & -3 & 1 & 3 & 0 & -5 & -3 & 4 & 5 & -2 & -4 & -2 & 3 & 1 & 0 & -1 \\
&&&&&&&&&&&&&&&&&&&&&&&&&& \\
0 & 0 & 0 & 0 & 1 & -1 & -1 & 2 & 2 & -2 & -6 & 4 & 10 & -1 & -12 & -6 & 12 & 11 & -5 & -10 & -2 & 7 & 3 & -2 & -2 & 0 & 1 \\
&&&&&&&&&&&&&&&&&&&&&&&&&& \\
0 & -1 & 2 & -1 & -2 & 1 & 5 & -1 & -12 & 0 & 18 & 7 & -18 & -20 & 14 & 22 & -1 & -18 & -7 & 10 & 6 & -3 & -3 & 0 & 2 & -1 & 0 \\
&&&&&&&&&&&&&&&&&&&&&&&&&& \\
1 & -1 & -2 & 1 & 5 & 1 & -11 & -5 & 14 & 13 & -9 & -23 & 2 & 20 & 8 & -11 & -11 & 4 & 6 & 0 & -2 & -1 & 1 & 0 & 0 & 0 & 0 \\
&&&&&&&&&&&&&&&&&&&&&&&&&& \\
-1 & 1 & 2 & 1 & -6 & -4 & 7 & 9 & -2 & -14 & -2 & 9 & 7 & -4 & -6 & 1 & 2 & 1 & -1 & 0 & 0 & 0 & 0 & 0 & 0 & 0 & 0 \\
&&&&&&&&&&&&&&&&&&&&&&&&&& \\
0 & 1 & -2 & -1 & 2 & 3 & 0 & -6 & 0 & 3 & 2 & -1 & -2 & 1 & 0 & 0 & 0 & 0 & 0 & 0 & 0 & 0 & 0 & 0 & 0 & 0 & 0
\end{array}\right).
\end{equation}
\begin{equation}
\mathcal{H}_{[2,1],[2,1]}=\frac{1}{q^{20}A^{9}\{A\}\{Aq\}\{A/q\}}
\left(\begin{array}{rrrrrrrrrrrrrrrrrrrrr}
0 & 0 & 0 & 0 & 0 & 1 & -2 & 3 & -4 & 5 & -5 & 5 & -4 & 3 & -2 & 1 & 0 & 0 & 0 & 0 & 0 \\
&&&&&&&&&&&&&&&&&&&& \\
0 & 0 & -1 & 2 & -4 & 7 & -11 & 14 & -18 & 20 & -21 & 20 & -18 & 14 & -11 & 7 & -4 & 2 & -1 & 0 & 0 \\
&&&&&&&&&&&&&&&&&&&& \\
1 & -3 & 8 & -17 & 29 & -43 & 63 & -81 & 94 & -105 & 111 & -105 & 94 & -81 & 63 & -43 & 29 & -17 & 8 & -3 & 1 \\
&&&&&&&&&&&&&&&&&&&& \\
-2 & 7 & -15 & 29 & -52 & 77 & -103 & 133 & -160 & 174 & -177 & 174 & -160 & 133 & -103 & 77 & -52 & 29 & -15 & 7 & -2 \\
&&&&&&&&&&&&&&&&&&&& \\
1 & -4 & 9 & -16 & 28 & -45 & 61 & -77 & 95 & -105 & 106 & -105 & 95 & -77 & 61 & -45 & 28 & -16 & 9 & -4 & 1 \\
&&&&&&&&&&&&&&&&&&&& \\
0 & 0 & -1 & 3 & -3 & 3 & -8 & 10 & -5 & 8 & -14 & 8 & -5 & 10 & -8 & 3 & -3 & 3 & -1 & 0 & 0 \\
&&&&&&&&&&&&&&&&&&&& \\
0 & 0 & 0 & 0 & 0 & 1 & -4 & 6 & -6 & 9 & -12 & 9 & -6 & 6 & -4 & 1 & 0 & 0 & 0 & 0 & 0
\end{array}\right).
\end{equation}
\end{landscape}
\newpage

\subsection{Borromean rings\label{A.b}}

\begin{equation}
\mathcal{H}_{[1],[1],[1]}=\frac{1}{q^{6}A^{2}\{A\}^2}
\left(\begin{array}{rrrrrrr}
0 & -1 & 4 & -5 & 4 & -1 & 0 \\
&&&&&& \\
1 & -4 & 7 & -10 & 7 & -4 & 1 \\
&&&&&& \\
0 & -1 & 4 & -5 & 4 & -1 & 0
\end{array}\right).
\end{equation}
\begin{equation}
\mathcal{H}_{[1],[1],[2]}=\frac{1}{q^{8}A^{2}\{A\}^2}
\left(\begin{array}{rrrrrrrrr}
0 & 0 & -1 & 3 & -1 & -2 & 3 & -1 & 0 \\
&&&&&&&& \\
1 & -3 & 2 & 3 & -8 & 3 & 2 & -3 & 1 \\
&&&&&&&& \\
0 & -1 & 3 & -2 & -1 & 3 & -1 & 0 & 0
\end{array}\right).
\end{equation}
\begin{equation}
\mathcal{H}_{[1],[2],[2]}=\frac{1}{q^{11}A^{3}\{A\}^2\{Aq\}}
\left(\begin{array}{rrrrrrrrrrrr}
0 & 0 & 0 & 0 & -1 & 2 & 2 & -4 & 1 & 2 & -1 & 0 \\
&&&&&&&&&&& \\
0 & 0 & 2 & -4 & -2 & 8 & -6 & -5 & 6 & -1 & -2 & 1 \\
&&&&&&&&&&& \\
-1 & 2 & 1 & -6 & 5 & 6 & -8 & 2 & 4 & -2 & 0 & 0 \\
&&&&&&&&&&& \\
0 & 1 & -2 & -1 & 4 & -2 & -2 & 1 & 0 & 0 & 0 & 0
\end{array}\right).
\end{equation}
\begin{equation}
\mathcal{H}_{[2],[2],[1,1]}=\frac{1}{q^{13}A^{3}\{A\}^2\{Aq\}}
\left(\begin{array}{rrrrrrrrrrrrrr}
0 & 0 & 0 & 0 & -1 & 1 & 3 & -2 & -3 & 3 & 1 & -1 & 0 & 0 \\
&&&&&&&&&&&&& \\
0 & 0 & 2 & -2 & -6 & 6 & 6 & -9 & -5 & 5 & 3 & -3 & -1 & 1 \\
&&&&&&&&&&&&& \\
-1 & 1 & 3 & -3 & -5 & 5 & 9 & -6 & -6 & 6 & 2 & -2 & 0 & 0 \\
&&&&&&&&&&&&& \\
0 & 0 & 1 & -1 & -3 & 3 & 2 & -3 & -1 & 1 & 0 & 0 & 0 & 0
\end{array}\right).
\end{equation}
\begin{equation}
\mathcal{H}_{[1],[1],[3]}=\frac{1}{q^{10}A^{2}\{A\}^2}
\left(\begin{array}{rrrrrrrrrrr}
0 & 0 & 0 & -1 & 3 & -2 & 2 & -3 & 3 & -1 & 0 \\
&&&&&&&&&& \\
1 & -3 & 3 & -2 & 4 & -8 & 4 & -2 & 3 & -3 & 1 \\
&&&&&&&&&& \\
0 & -1 & 3 & -3 & 2 & -2 & 3 & -1 & 0 & 0 & 0
\end{array}\right).
\end{equation}
\begin{equation}
\mathcal{H}_{[1],[3],[3]}=\frac{1}{q^{15}A^{3}\{A\}^2\{Aq\}}
\left(\begin{array}{rrrrrrrrrrrrrrrr}
0 & 0 & 0 & 0 & 0 & 0 & -1 & 2 & 0 & 2 & -4 & 2 & -1 & 2 & -1 & 0 \\
&&&&&&&&&&&&&&& \\
0 & 0 & 0 & 2 & -4 & 2 & -4 & 8 & -8 & 3 & -6 & 6 & -2 & 1 & -2 & 1 \\
&&&&&&&&&&&&&&& \\
-1 & 2 & -1 & 2 & -6 & 6 & -3 & 8 & -8 & 4 & -2 & 4 & -2 & 0 & 0 & 0 \\
&&&&&&&&&&&&&&& \\
0 & 1 & -2 & 1 & -2 & 4 & -2 & 0 & -2 & 1 & 0 & 0 & 0 & 0 & 0 & 0
\end{array}\right).
\end{equation}
\begin{equation}
\mathcal{H}_{[1],[1],[2,1]}=\frac{1}{q^{10}A^{2}\{A\}^2}
\left(\begin{array}{rrrrrrrrrrr}
0 & 0 & -1 & 3 & -3 & 3 & -3 & 3 & -1 & 0 & 0 \\
&&&&&&&&&& \\
1 & -4 & 8 & -12 & 15 & -18 & 15 & -12 & 8 & -4 & 1 \\
&&&&&&&&&& \\
0 & 0 & -1 & 3 & -3 & 3 & -3 & 3 & -1 & 0 & 0
\end{array}\right).
\end{equation}
\begin{equation}
\mathcal{H}_{[1],[1,1,1],[3]}=\frac{1}{q^{14}A^{2}\{A\}^2}
\left(\begin{array}{rrrrrrrrrrrrrrr}
0 & 0 & 0 & -1 & 2 & -1 & 2 & -3 & 2 & -1 & 2 & -1 & 0 & 0 & 0 \\
&&&&&&&&&&&&&& \\
1 & -2 & 1 & -2 & 4 & -2 & 2 & -6 & 2 & -2 & 4 & -2 & 1 & -2 & 1 \\
&&&&&&&&&&&&&& \\
0 & 0 & 0 & -1 & 2 & -1 & 2 & -3 & 2 & -1 & 2 & -1 & 0 & 0 & 0
\end{array}\right).
\end{equation}
\begin{equation}
\mathcal{H}_{[1,1],[1,1],[3]}=\frac{1}{q^{15}A^{3}\{A\}^2\{Aq^{-1}\}}
\left(\begin{array}{rrrrrrrrrrrrrrrr}
0 & 0 & 0 & -1 & 1 & 2 & -1 & -1 & -1 & 2 & 1 & -1 & 0 & 0 & 0 & 0 \\
&&&&&&&&&&&&&&& \\
1 & -1 & -2 & 1 & 2 & 3 & -4 & -7 & 2 & 4 & 2 & -4 & -2 & 2 & 0 & 0 \\
&&&&&&&&&&&&&&& \\
0 & 0 & -2 & 2 & 4 & -2 & -4 & -2 & 7 & 4 & -3 & -2 & -1 & 2 & 1 & -1 \\
&&&&&&&&&&&&&&& \\
0 & 0 & 0 & 0 & 1 & -1 & -2 & 1 & 1 & 1 & -2 & -1 & 1 & 0 & 0 & 0
\end{array}\right).
\end{equation}
\begin{equation}
\mathcal{H}_{[1,1],[3],[3]}=\frac{1}{q^{17}A^{3}\{A\}^2\{Aq\}}
\left(\begin{array}{rrrrrrrrrrrrrrrrrr}
0 & 0 & 0 & 0 & 0 & 0 & -1 & 1 & 1 & 2 & -2 & -2 & 1 & 1 & 1 & -1 & 0 & 0 \\
&&&&&&&&&&&&&&&&& \\
0 & 0 & 0 & 2 & -2 & -2 & -2 & 4 & 4 & -5 & -3 & -3 & 4 & 2 & -1 & -1 & -1 & 1 \\
&&&&&&&&&&&&&&&&& \\
-1 & 1 & 1 & 1 & -2 & -4 & 3 & 3 & 5 & -4 & -4 & 2 & 2 & 2 & -2 & 0 & 0 & 0 \\
&&&&&&&&&&&&&&&&& \\
0 & 0 & 1 & -1 & -1 & -1 & 2 & 2 & -2 & -1 & -1 & 1 & 0 & 0 & 0 & 0 & 0 & 0
\end{array}\right).
\end{equation}
\begin{equation*}
\mathcal{H}_{[2],[2],[2,1]}=\frac{1}{q^{20}A^{4}\{A\}^2\{Aq\}^2}
\left(\begin{array}{rrrrrrrrrrrrrrrrrrrrr}
0 & 0 & 0 & 0 & 0 & 1 & -4 & 3 & 6 & -9 & -1 & 6 & 4 & -4 & -8 & 8 & 2 & -4 & 1 & 0 & 0 \\
&&&&&&&&&&&&&&&&&&&& \\
0 & 0 & -1 & 3 & 1 & -11 & 10 & 5 & -9 & -3 & -6 & 19 & -6 & -13 & 6 & -1 & 7 & -5 & -3 & 4 & -1 \\
&&&&&&&&&&&&&&&&&&&& \\
1 & -4 & 4 & 4 & -13 & 9 & 2 & 3 & -8 & -11 & 32 & -11 & -8 & 3 & 2 & 9 & -13 & 4 & 4 & -4 & 1 \\
&&&&&&&&&&&&&&&&&&&& \\
-1 & 4 & -3 & -5 & 7 & -1 & 6 & -13 & -6 & 19 & -6 & -3 & -9 & 5 & 10 & -11 & 1 & 3 & -1 & 0 & 0 \\
&&&&&&&&&&&&&&&&&&&& \\
0 & 0 & 1 & -4 & 2 & 8 & -8 & -4 & 4 & 6 & -1 & -9 & 6 & 3 & -4 & 1 & 0 & 0 & 0 & 0 & 0
\end{array}\right).
\end{equation*}
\begin{equation}
\mathcal{H}_{[2],[1,1,1],[3]}=\frac{1}{q^{8}A^{1}\{A\}^2\{Aq\}}
\left(\begin{array}{rrrrrrrrrrr}
0 & 0 & 1 & 0 & -1 & 0 & 0 & 1 & 0 & 0 & 0 \\
&&&&&&&&&& \\
-1 & 0 & 1 & 1 & 0 & -3 & 0 & 1 & 1 & 0 & -1 \\
&&&&&&&&&& \\
0 & 0 & 0 & 1 & 0 & -1 & -1 & 0 & 1 & 0 & 0
\end{array}\right).
\end{equation}
\begin{equation}
\mathcal{H}_{[1,1,1],[3],[3]}=\frac{1}{q^{19}A^{3}\{A\}^2\{Aq\}}
\left(\begin{array}{rrrrrrrrrrrrrrrrrrrr}
0 & 0 & 0 & 0 & 0 & 0 & -1 & 1 & 0 & 3 & -2 & 0 & -3 & 3 & 0 & 1 & -1 & 0 & 0 & 0 \\
&&&&&&&&&&&&&&&&&&& \\
0 & 0 & 0 & 2 & -2 & 0 & -6 & 6 & 0 & 6 & -9 & 0 & -5 & 5 & 0 & 3 & -3 & 0 & -1 & 1 \\
&&&&&&&&&&&&&&&&&&& \\
-1 & 1 & 0 & 3 & -3 & 0 & -5 & 5 & 0 & 9 & -6 & 0 & -6 & 6 & 0 & 2 & -2 & 0 & 0 & 0 \\
&&&&&&&&&&&&&&&&&&& \\
0 & 0 & 0 & 1 & -1 & 0 & -3 & 3 & 0 & 2 & -3 & 0 & -1 & 1 & 0 & 0 & 0 & 0 & 0 & 0
\end{array}\right).
\end{equation}

\newpage

\begin{landscape}

\begin{equation}
\mathcal{H}_{[2],[2],[2]}=\frac{1}{q^{18}A^{4}\{A\}^2\{Aq\}^2}
\left(\begin{array}{rrrrrrrrrrrrrrrrrrr}
0 & 0 & 0 & 0 & 0 & 1 & -4 & 2 & 11 & -16 & -4 & 25 & -11 & -14 & 13 & 1 & -4 & 1 & 0 \\
&&&&&&&&&&&&&&&&&& \\
0 & 0 & -1 & 3 & 2 & -15 & 11 & 24 & -43 & -7 & 56 & -27 & -34 & 33 & 5 & -16 & 3 & 3 & -1 \\
&&&&&&&&&&&&&&&&&& \\
1 & -4 & 3 & 10 & -25 & 7 & 47 & -54 & -25 & 86 & -25 & -54 & 47 & 7 & -25 & 10 & 3 & -4 & 1 \\
&&&&&&&&&&&&&&&&&& \\
-1 & 3 & 3 & -16 & 5 & 33 & -34 & -27 & 56 & -7 & -43 & 24 & 11 & -15 & 2 & 3 & -1 & 0 & 0 \\
&&&&&&&&&&&&&&&&&& \\
0 & 1 & -4 & 1 & 13 & -14 & -11 & 25 & -4 & -16 & 11 & 2 & -4 & 1 & 0 & 0 & 0 & 0 & 0
\end{array}\right).
\end{equation}
\begin{equation}
\mathcal{H}_{[2],[2],[3]}=\frac{1}{q^{22}A^{4}\{A\}^2\{Aq\}^2}
\left(\begin{array}{rrrrrrrrrrrrrrrrrrrrrrr}
0 & 0 & 0 & 0 & 0 & 0 & 0 & 1 & -3 & -1 & 9 & -2 & -10 & 2 & 10 & 2 & -11 & -2 & 9 & -1 & -3 & 1 & 0 \\
&&&&&&&&&&&&&&&&&&&&&& \\
0 & 0 & 0 & -1 & 2 & 4 & -9 & -7 & 19 & 10 & -27 & -19 & 28 & 24 & -26 & -23 & 17 & 16 & -9 & -8 & 4 & 2 & -1 \\
&&&&&&&&&&&&&&&&&&&&&& \\
1 & -3 & -1 & 11 & -4 & -21 & 9 & 33 & -6 & -48 & 2 & 60 & 2 & -48 & -6 & 33 & 9 & -21 & -4 & 11 & -1 & -3 & 1 \\
&&&&&&&&&&&&&&&&&&&&&& \\
-1 & 2 & 4 & -8 & -9 & 16 & 17 & -23 & -26 & 24 & 28 & -19 & -27 & 10 & 19 & -7 & -9 & 4 & 2 & -1 & 0 & 0 & 0 \\
&&&&&&&&&&&&&&&&&&&&&& \\
0 & 1 & -3 & -1 & 9 & -2 & -11 & 2 & 10 & 2 & -10 & -2 & 9 & -1 & -3 & 1 & 0 & 0 & 0 & 0 & 0 & 0 & 0
\end{array}\right).
\end{equation}
\begin{equation*}
\mathcal{H}_{[2],[3],[3]}=\frac{1}{q^{28}A^{5}\{A\}^2\{Aq\}^2\{Aq\}}
\left(\begin{array}{rrrrrrrrrrrrrrrrrrrrrrrrrrrrr}
0 & 0 & 0 & 0 & 0 & 0 & 0 & 0 & 0 & 0 & 0 & 1 & -2 & -3 & 5 & 7 & -4 & -12 & 0 & 16 & 3 & -12 & -5 & 5 & 6 & -3 & -2 & 1 & 0 \\
&&&&&&&&&&&&&&&&&&&&&&&&&&&& \\
0 & 0 & 0 & 0 & 0 & 0 & 0 & -2 & 3 & 8 & -9 & -17 & 7 & 33 & 1 & -49 & -17 & 44 & 31 & -28 & -36 & 12 & 25 & -1 & -12 & -3 & 5 & 1 & -1 \\
&&&&&&&&&&&&&&&&&&&&&&&&&&&& \\
0 & 0 & 0 & 1 & 0 & -7 & 0 & 20 & 4 & -39 & -19 & 55 & 52 & -56 & -78 & 40 & 89 & -3 & -75 & -17 & 48 & 20 & -22 & -13 & 7 & 7 & -3 & -2 & 1 \\
&&&&&&&&&&&&&&&&&&&&&&&&&&&& \\
-1 & 2 & 3 & -7 & -7 & 13 & 22 & -20 & -48 & 17 & 75 & 3 & -89 & -40 & 78 & 56 & -52 & -55 & 19 & 39 & -4 & -20 & 0 & 7 & 0 & -1 & 0 & 0 & 0 \\
&&&&&&&&&&&&&&&&&&&&&&&&&&&& \\
1 & -1 & -5 & 3 & 12 & 1 & -25 & -12 & 36 & 28 & -31 & -44 & 17 & 49 & -1 & -33 & -7 & 17 & 9 & -8 & -3 & 2 & 0 & 0 & 0 & 0 & 0 & 0 & 0 \\
&&&&&&&&&&&&&&&&&&&&&&&&&&&& \\
0 & -1 & 2 & 3 & -6 & -5 & 5 & 12 & -3 & -16 & 0 & 12 & 4 & -7 & -5 & 3 & 2 & -1 & 0 & 0 & 0 & 0 & 0 & 0 & 0 & 0 & 0 & 0 & 0
\end{array}\right).
\end{equation*}
\newpage
{\tiny{
\begin{equation*}
\mathcal{H}_{[3],[3],[3]}=\frac{1}{q^{38}A^{6}\{A\}^2\{Aq\}^2\{Aq^2\}^2}
\left(\begin{array}{rrrrrrrrrrrrrrrrrrrrrrrrrrrrrrrrrrrrrrr}
0 & 0 & 0 & 0 & 0 & 0 & 0 & 0 & 0 & 0 & 0 & 0 & 0 & -1 & 4 & -2 & -8 & 2 & 15 & 10 & -36 & -17 & 39 & 42 & -28 & -67 & 11 & 62 & 15 & -39 & -31 & 23 & 17 & -3 & -9 & -1 & 4 & -1 & 0 \\
&&&&&&&&&&&&&&&&&&&&&&&&&&&&&&&&&&&&&& \\
0 & 0 & 0 & 0 & 0 & 0 & 0 & 0 & 1 & -3 & -1 & 6 & 10 & -15 & -30 & 25 & 59 & -4 & -116 & -34 & 135 & 109 & -112 & -179 & 41 & 184 & 42 & -128 & -91 & 63 & 71 & -4 & -43 & -9 & 15 & 6 & -2 & -3 & 1 \\
&&&&&&&&&&&&&&&&&&&&&&&&&&&&&&&&&&&&&& \\
0 & 0 & 0 & 0 & -1 & 3 & 1 & -7 & -7 & 15 & 29 & -35 & -65 & 36 & 135 & 7 & -224 & -97 & 253 & 243 & -172 & -357 & 42 & 340 & 113 & -232 & -179 & 109 & 146 & -15 & -87 & -19 & 41 & 14 & -11 & -8 & 2 & 3 & -1 \\
&&&&&&&&&&&&&&&&&&&&&&&&&&&&&&&&&&&&&& \\
0 & 1 & -4 & 3 & 6 & -5 & -16 & 8 & 51 & -22 & -100 & -3 & 185 & 87 & -259 & -236 & 238 & 389 & -89 & -488 & -89 & 389 & 238 & -236 & -259 & 87 & 185 & -3 & -100 & -22 & 51 & 8 & -16 & -5 & 6 & 3 & -4 & 1 & 0 \\
&&&&&&&&&&&&&&&&&&&&&&&&&&&&&&&&&&&&&& \\
-1 & 3 & 2 & -8 & -11 & 14 & 41 & -19 & -87 & -15 & 146 & 109 & -179 & -232 & 113 & 340 & 42 & -357 & -172 & 243 & 253 & -97 & -224 & 7 & 135 & 36 & -65 & -35 & 29 & 15 & -7 & -7 & 1 & 3 & -1 & 0 & 0 & 0 & 0 \\
&&&&&&&&&&&&&&&&&&&&&&&&&&&&&&&&&&&&&& \\
1 & -3 & -2 & 6 & 15 & -9 & -43 & -4 & 71 & 63 & -91 & -128 & 42 & 184 & 41 & -179 & -112 & 109 & 135 & -34 & -116 & -4 & 59 & 25 & -30 & -15 & 10 & 6 & -1 & -3 & 1 & 0 & 0 & 0 & 0 & 0 & 0 & 0 & 0 \\
&&&&&&&&&&&&&&&&&&&&&&&&&&&&&&&&&&&&&& \\
0 & -1 & 4 & -1 & -9 & -3 & 17 & 23 & -31 & -39 & 15 & 62 & 11 & -67 & -28 & 42 & 39 & -17 & -36 & 10 & 15 & 2 & -8 & -2 & 4 & -1 & 0 & 0 & 0 & 0 & 0 & 0 & 0 & 0 & 0 & 0 & 0 & 0 & 0
\end{array}\right).
\end{equation*}
\begin{equation*}
\mathcal{H}_{[2,1],[2,1],[2,1]}=\frac{1}{q^{30}A^{6}\{A\}^2\{Aq\}^2\{A/q\}^2}
\left(\begin{array}{rrrrrrrrrrrrrrrrrrrrrrrrrrrrrrr}
0 & 0 & 0 & 0 & 0 & -1 & 8 & -27 & 57 & -101 & 169 & -253 & 332 & -399 & 454 & -477 & 454 & -399 & 332 & -253 & 169 & -101 & 57 & -27 & 8 & -1 & 0 & 0 & 0 & 0 & 0 \\
&&&&&&&&&&&&&&&&&&&&&&&&&&&&&& \\
0 & 0 & 1 & -7 & 20 & -37 & 67 & -124 & 200 & -283 & 391 & -527 & 641 & -724 & 798 & -838 & 798 & -724 & 641 & -527 & 391 & -283 & 200 & -124 & 67 & -37 & 20 & -7 & 1 & 0 & 0 \\
&&&&&&&&&&&&&&&&&&&&&&&&&&&&&& \\
-1 & 8 & -30 & 80 & -181 & 363 & -662 & 1110 & -1723 & 2497 & -3403 & 4370 & -5285 & 6054 & -6563 & 6747 & -6563 & 6054 & -5285 & 4370 & -3403 & 2497 & -1723 & 1110 & -662 & 363 & -181 & 80 & -30 & 8 & -1 \\
&&&&&&&&&&&&&&&&&&&&&&&&&&&&&& \\
2 & -16 & 58 & -146 & 323 & -654 & 1180 & -1928 & 2957 & -4262 & 5725 & -7246 & 8714 & -9942 & 10717 & -10984 & 10717 & -9942 & 8714 & -7246 & 5725 & -4262 & 2957 & -1928 & 1180 & -654 & 323 & -146 & 58 & -16 & 2 \\
&&&&&&&&&&&&&&&&&&&&&&&&&&&&&& \\
-1 & 8 & -30 & 80 & -181 & 363 & -662 & 1110 & -1723 & 2497 & -3403 & 4370 & -5285 & 6054 & -6563 & 6747 & -6563 & 6054 & -5285 & 4370 & -3403 & 2497 & -1723 & 1110 & -662 & 363 & -181 & 80 & -30 & 8 & -1 \\
&&&&&&&&&&&&&&&&&&&&&&&&&&&&&& \\
0 & 0 & 1 & -7 & 20 & -37 & 67 & -124 & 200 & -283 & 391 & -527 & 641 & -724 & 798 & -838 & 798 & -724 & 641 & -527 & 391 & -283 & 200 & -124 & 67 & -37 & 20 & -7 & 1 & 0 & 0 \\
&&&&&&&&&&&&&&&&&&&&&&&&&&&&&& \\
0 & 0 & 0 & 0 & 0 & -1 & 8 & -27 & 57 & -101 & 169 & -253 & 332 & -399 & 454 & -477 & 454 & -399 & 332 & -253 & 169 & -101 & 57 & -27 & 8 & -1 & 0 & 0 & 0 & 0 & 0
\end{array}\right).
\end{equation*}
\begin{equation*}
\mathcal{H}_{[3],[3],[2,1]}=\frac{1}{q^{30}A^{5}\{A\}^2\{Aq\}^2\{Aq^2\}}
\left(\begin{array}{rrrrrrrrrrrrrrrrrrrrrrrrrrrrrrr}
0 & 0 & 0 & 0 & 0 & 0 & 0 & 0 & 0 & 0 & 0 & 1 & -2 & -2 & 2 & 7 & 0 & -9 & -7 & 11 & 8 & -1 & -13 & 1 & 5 & 3 & -2 & -2 & 1 & 0 & 0 \\
&&&&&&&&&&&&&&&&&&&&&&&&&&&&&& \\
0 & 0 & 0 & 0 & 0 & 0 & 0 & -2 & 3 & 6 & -4 & -14 & -4 & 24 & 17 & -26 & -27 & -2 & 43 & 0 & -9 & -28 & 12 & 4 & 10 & -6 & -3 & -1 & 1 & 2 & -1 \\
&&&&&&&&&&&&&&&&&&&&&&&&&&&&&& \\
0 & 0 & 0 & 1 & 0 & -6 & -1 & 14 & 8 & -20 & -24 & 11 & 55 & -8 & -28 & -49 & 43 & 35 & 13 & -37 & -24 & 15 & 23 & 3 & -16 & -5 & 6 & 4 & -2 & -2 & 1 \\
&&&&&&&&&&&&&&&&&&&&&&&&&&&&&& \\
-1 & 2 & 2 & -4 & -6 & 5 & 16 & -3 & -23 & -15 & 24 & 37 & -13 & -35 & -43 & 49 & 28 & 8 & -55 & -11 & 24 & 20 & -8 & -14 & 1 & 6 & 0 & -1 & 0 & 0 & 0 \\
&&&&&&&&&&&&&&&&&&&&&&&&&&&&&& \\
1 & -2 & -1 & 1 & 3 & 6 & -10 & -4 & -12 & 28 & 9 & 0 & -43 & 2 & 27 & 26 & -17 & -24 & 4 & 14 & 4 & -6 & -3 & 2 & 0 & 0 & 0 & 0 & 0 & 0 & 0 \\
&&&&&&&&&&&&&&&&&&&&&&&&&&&&&& \\
0 & 0 & -1 & 2 & 2 & -3 & -5 & -1 & 13 & 1 & -8 & -11 & 7 & 9 & 0 & -7 & -2 & 2 & 2 & -1 & 0 & 0 & 0 & 0 & 0 & 0 & 0 & 0 & 0 & 0 & 0
\end{array}\right).
\end{equation*}
}}
\end{landscape}
\newpage

\subsection{L7a3 link\label{A.l}}

Since link L7a3 is asymmetric in two components, it is important to fix the order of the components. Here the first component is the unknot and the second one is the trefoil.

\begin{equation}
\mathcal{H}_{[1],[1]}=\frac{1}{q^{6}A^{2}\{A\}}
\left(\begin{array}{rrrrrrr}
0 & 1 & -1 & 2 & -1 & 1 & 0 \\
&&&&&& \\
-1 & 2 & -4 & 3 & -4 & 2 & -1 \\
&&&&&& \\
0 & 1 & -2 & 3 & -2 & 1 & 0
\end{array}\right).
\end{equation}
\begin{equation}
\mathcal{H}_{[1],[2]}=\frac{1}{q^{8}A^{3}\{A\}}
\left(\begin{array}{rrrrrrrrrrrrr}
0 & 0 & 1 & 0 & 0 & 2 & -1 & 0 & 2 & 0 & -1 & 1 & 0 \\
&&&&&&&&&&&& \\
-1 & 1 & -1 & -3 & 3 & -3 & -4 & 3 & 0 & -4 & 1 & 1 & -1 \\
&&&&&&&&&&&& \\
1 & 0 & -2 & 4 & 1 & -4 & 4 & 2 & -2 & 0 & 1 & 0 & 0 \\
&&&&&&&&&&&& \\
0 & -1 & 1 & 1 & -3 & 1 & 1 & -1 & 0 & 0 & 0 & 0 & 0
\end{array}\right).
\end{equation}
\begin{equation}
\mathcal{H}_{[2],[1]}=\frac{1}{q^{8}A^{2}\{A\}}
\left(\begin{array}{rrrrrrrrr}
0 & 0 & 1 & 0 & 0 & 1 & -1 & 1 & 0 \\
&&&&&&&& \\
-1 & 1 & 0 & -2 & 1 & -2 & 0 & 1 & -1 \\
&&&&&&&& \\
0 & 1 & -1 & 0 & 1 & -1 & 1 & 0 & 0
\end{array}\right).
\end{equation}
\begin{equation}
\mathcal{H}_{[2],[2]}=\frac{1}{q^{19}A^{4}\{A\}\{Aq\}}
\left(\begin{array}{rrrrrrrrrrrrrrrr}
0 & 0 & 1 & 0 & -1 & 1 & 2 & -1 & 0 & 1 & 0 & 0 & 1 & 0 & 0 & 0 \\
&&&&&&&&&&&&&&& \\
-1 & 0 & 2 & -2 & -4 & 2 & 1 & -4 & -2 & 0 & -1 & -1 & 0 & -1 & -1 & 0 \\
&&&&&&&&&&&&&&& \\
0 & 1 & 1 & -1 & -1 & 3 & 3 & -1 & -1 & 4 & 2 & 1 & 0 & 0 & 1 & 1 \\
&&&&&&&&&&&&&&& \\
0 & 0 & 0 & -1 & -1 & 1 & 1 & -2 & -2 & 1 & 1 & -2 & -2 & 1 & 0 & -1 \\
&&&&&&&&&&&&&&& \\
0 & 0 & 0 & 0 & 0 & 0 & 1 & 0 & -2 & 1 & 2 & -2 & 0 & 1 & 0 & 0
\end{array}\right).
\end{equation}
\begin{equation}
\mathcal{H}_{[2],[1,1]}=\frac{1}{q^{18}A^{3}\{A\}}
\left(\begin{array}{rrrrrrrrrrrrrrr}
0 & 0 & 1 & 0 & -1 & 1 & 2 & -1 & 0 & 1 & 0 & 0 & 1 & 0 & 0 \\
&&&&&&&&&&&&&& \\
-1 & 0 & 2 & -1 & -4 & 1 & 2 & -2 & -3 & 0 & 0 & -1 & 0 & 0 & -1 \\
&&&&&&&&&&&&&& \\
0 & 0 & 1 & 1 & -2 & -1 & 4 & 1 & -3 & 1 & 2 & 1 & -1 & 0 & 1 \\
&&&&&&&&&&&&&& \\
0 & 0 & 0 & 0 & 0 & -1 & 0 & 2 & -1 & -2 & 2 & 0 & -1 & 0 & 0
\end{array}\right).
\end{equation}
\begin{equation}
\mathcal{H}_{[1],[3]}=\frac{1}{q^{10}A^{4}\{A\}}
\left(\begin{array}{rrrrrrrrrrrrrrrrrrrrrr}
0 & 0 & 0 & 0 & 1 & 0 & 1 & 0 & 2 & -1 & 1 & 1 & 2 & -1 & 0 & 0 & 2 & 0 & 0 & -1 & 1 & 0 \\
&&&&&&&&&&&&&&&&&&&&& \\
0 & -1 & 1 & -2 & 0 & -3 & 2 & -5 & -2 & -3 & 4 & -4 & -3 & -4 & 3 & 0 & 0 & -4 & 1 & 0 & 1 & -1 \\
&&&&&&&&&&&&&&&&&&&&& \\
1 & 0 & 0 & 0 & 4 & 1 & -1 & 0 & 6 & 4 & -1 & -3 & 3 & 3 & 2 & -2 & 0 & 0 & 1 & 0 & 0 & 0 \\
&&&&&&&&&&&&&&&&&&&&& \\
-1 & 0 & 0 & 2 & -4 & -1 & -1 & 4 & -3 & -2 & -2 & 2 & 0 & 0 & -1 & 0 & 0 & 0 & 0 & 0 & 0 & 0 \\
&&&&&&&&&&&&&&&&&&&&& \\
0 & 1 & -1 & 0 & -1 & 3 & -1 & 0 & -1 & 1 & 0 & 0 & 0 & 0 & 0 & 0 & 0 & 0 & 0 & 0 & 0 & 0
\end{array}\right).
\end{equation}
\begin{equation}
\mathcal{H}_{[3],[1]}=\frac{1}{q^{10}A^{2}\{A\}}
\left(\begin{array}{rrrrrrrrrrr}
0 & 0 & 0 & 1 & 0 & 1 & -1 & 1 & -1 & 1 & 0 \\
&&&&&&&&&& \\
-1 & 1 & -1 & 2 & -3 & 1 & -3 & 2 & -1 & 1 & -1 \\
&&&&&&&&&& \\
0 & 1 & -1 & 1 & -2 & 2 & -1 & 1 & 0 & 0 & 0
\end{array}\right).
\end{equation}
\begin{equation}
\mathcal{H}_{[2,1],[1]}=\frac{1}{q^{10}A^{2}\{A\}}
\left(\begin{array}{rrrrrrrrrrr}
0 & 0 & 1 & -1 & 2 & -2 & 2 & -1 & 1 & 0 & 0 \\
&&&&&&&&&& \\
-1 & 2 & -4 & 6 & -8 & 7 & -8 & 6 & -4 & 2 & -1 \\
&&&&&&&&&& \\
0 & 0 & 1 & -1 & 1 & -1 & 1 & -1 & 1 & 0 & 0
\end{array}\right).
\end{equation}
\begin{equation}
\mathcal{H}_{[2],[3]}=\frac{1}{q^{19}A^{4}\{A\}\{Aq\}}
\left(\begin{array}{rrrrrrrrrrrrrrrr}
0 & 0 & 1 & 0 & -1 & 1 & 2 & -1 & 0 & 1 & 0 & 0 & 1 & 0 & 0 & 0 \\
&&&&&&&&&&&&&&& \\
-1 & 0 & 2 & -2 & -4 & 2 & 1 & -4 & -2 & 0 & -1 & -1 & 0 & -1 & -1 & 0 \\
&&&&&&&&&&&&&&& \\
0 & 1 & 1 & -1 & -1 & 3 & 3 & -1 & -1 & 4 & 2 & 1 & 0 & 0 & 1 & 1 \\
&&&&&&&&&&&&&&& \\
0 & 0 & 0 & -1 & -1 & 1 & 1 & -2 & -2 & 1 & 1 & -2 & -2 & 1 & 0 & -1 \\
&&&&&&&&&&&&&&& \\
0 & 0 & 0 & 0 & 0 & 0 & 1 & 0 & -2 & 1 & 2 & -2 & 0 & 1 & 0 & 0
\end{array}\right).
\end{equation}
\begin{equation}
\mathcal{H}_{[2],[1,1,1]}=\frac{1}{q^{34}A^{4}\{A\}}
\left(\begin{array}{rrrrrrrrrrrrrrrrrrrrrrrr}
0 & 0 & 1 & 0 & -1 & 0 & 1 & 2 & 0 & -1 & 0 & 2 & 1 & 0 & 0 & 1 & 0 & 1 & 0 & 1 & 0 & 0 & 0 & 0 \\
&&&&&&&&&&&&&&&&&&&&&&& \\
-1 & 0 & 1 & 1 & -1 & -4 & -1 & 2 & 2 & -2 & -5 & -3 & 1 & 1 & -2 & -4 & -1 & -1 & 0 & -1 & -1 & 0 & -1 & 0 \\
&&&&&&&&&&&&&&&&&&&&&&& \\
0 & 0 & 0 & 1 & 1 & 0 & -2 & -1 & 3 & 5 & 1 & -4 & -1 & 4 & 5 & 1 & -1 & 0 & 3 & 1 & 1 & -1 & 1 & 1 \\
&&&&&&&&&&&&&&&&&&&&&&& \\
0 & 0 & 0 & 0 & 0 & 0 & 0 & -1 & -1 & 0 & 2 & 1 & -3 & -3 & 0 & 3 & 0 & -2 & -2 & 0 & 0 & 0 & 0 & -1 \\
&&&&&&&&&&&&&&&&&&&&&&& \\
0 & 0 & 0 & 0 & 0 & 0 & 0 & 0 & 0 & 0 & 0 & 0 & 1 & 0 & -1 & -1 & 1 & 2 & -1 & -1 & 0 & 1 & 0 & 0
\end{array}\right).
\end{equation}
\begin{equation}
\mathcal{H}_{[3],[2]}=\frac{1}{q^{19}A^{4}\{A\}\{Aq\}}
\left(\begin{array}{rrrrrrrrrrrrrrrrrrrrrrr}
0 & 0 & 0 & 0 & 0 & 0 & 0 & 1 & 0 & 0 & 1 & 0 & 0 & 1 & 1 & 0 & -2 & 1 & 2 & -1 & -1 & 1 & 0 \\
&&&&&&&&&&&&&&&&&&&&&& \\
0 & 0 & 0 & -1 & 0 & 1 & -1 & -1 & -1 & -2 & -2 & 0 & 2 & -3 & -5 & 1 & 4 & -1 & -4 & 0 & 2 & 0 & -1 \\
&&&&&&&&&&&&&&&&&&&&&& \\
1 & -1 & -1 & 3 & 0 & -1 & 2 & 0 & -2 & 1 & 9 & 3 & -9 & -2 & 11 & 3 & -6 & -3 & 4 & 3 & -2 & -1 & 1 \\
&&&&&&&&&&&&&&&&&&&&&& \\
-1 & 0 & 2 & -1 & -2 & 2 & -1 & -4 & 2 & 6 & -3 & -10 & 1 & 9 & -1 & -7 & 0 & 3 & 0 & -1 & 0 & 0 & 0 \\
&&&&&&&&&&&&&&&&&&&&&& \\
0 & 1 & -1 & -1 & 2 & -1 & -1 & 2 & 1 & -1 & -3 & 2 & 3 & -2 & -1 & 1 & 0 & 0 & 0 & 0 & 0 & 0 & 0
\end{array}\right).
\end{equation}
\begin{equation}
\mathcal{H}_{[1,1,1],[2]}=\frac{1}{q^{12}A^{3}\{A\}}
\left(\begin{array}{rrrrrrrrrrrrrrrrr}
0 & 0 & 1 & 0 & 0 & 1 & -1 & 0 & 1 & 1 & 0 & 0 & 0 & 1 & 0 & 0 & 0 \\
&&&&&&&&&&&&&&&& \\
-1 & 0 & -1 & 0 & 2 & -2 & -1 & -2 & -1 & 1 & 0 & -2 & -1 & 0 & 1 & 0 & -1 \\
&&&&&&&&&&&&&&&& \\
1 & 0 & -1 & 1 & 0 & 2 & 1 & -2 & 1 & 1 & 1 & -1 & -1 & 1 & 1 & 0 & 0 \\
&&&&&&&&&&&&&&&& \\
0 & 0 & 0 & -1 & 0 & 1 & -1 & 0 & 0 & 1 & 0 & -1 & 0 & 0 & 0 & 0 & 0
\end{array}\right).
\end{equation}
\begin{equation}
\mathcal{H}_{[3],[1,1,1]}=\frac{1}{q^{19}A^{4}\{A\}}
\left(\begin{array}{rrrrrrrrrrrrrrrr}
0 & 0 & 1 & 0 & -1 & 1 & 2 & -1 & 0 & 1 & 0 & 0 & 1 & 0 & 0 & 0 \\
&&&&&&&&&&&&&&& \\
-1 & 0 & 2 & -2 & -4 & 2 & 1 & -4 & -2 & 0 & -1 & -1 & 0 & -1 & -1 & 0 \\
&&&&&&&&&&&&&&& \\
0 & 1 & 1 & -1 & -1 & 3 & 3 & -1 & -1 & 4 & 2 & 1 & 0 & 0 & 1 & 1 \\
&&&&&&&&&&&&&&& \\
0 & 0 & 0 & -1 & -1 & 1 & 1 & -2 & -2 & 1 & 1 & -2 & -2 & 1 & 0 & -1 \\
&&&&&&&&&&&&&&& \\
0 & 0 & 0 & 0 & 0 & 0 & 1 & 0 & -2 & 1 & 2 & -2 & 0 & 1 & 0 & 0
\end{array}\right).
\end{equation}

\newpage

\begin{landscape}
{\tiny{
\begin{equation}
\mathcal{H}_{[3],[3]}=\frac{1}{q^{29}A^{6}\{A\}\{Aq\}\{Aq^2\}}
\left(\begin{array}{rrrrrrrrrrrrrrrrrrrrrrrrrrrrrrrrrrrrrrr}
0 & 0 & 0 & 0 & 0 & 0 & 0 & 0 & 0 & 0 & 0 & 0 & 0 & 1 & -1 & 1 & 1 & 1 & -1 & 1 & 2 & 0 & -2 & 2 & 3 & 1 & -5 & 0 & 4 & 3 & -3 & -3 & 1 & 3 & 0 & -1 & -1 & 1 & 0 \\
&&&&&&&&&&&&&&&&&&&&&&&&&&&&&&&&&&&&&& \\
0 & 0 & 0 & 0 & 0 & 0 & 0 & 0 & -1 & 1 & -1 & 0 & -3 & 1 & -2 & -2 & -4 & 0 & -4 & -5 & -4 & 5 & -1 & -7 & -11 & 5 & 8 & 1 & -12 & -6 & 5 & 8 & -2 & -5 & -3 & 3 & 1 & 0 & -1 \\
&&&&&&&&&&&&&&&&&&&&&&&&&&&&&&&&&&&&&& \\
0 & 0 & 0 & 0 & 1 & -1 & 0 & 2 & 1 & 0 & -1 & 4 & 2 & 1 & 7 & 7 & -1 & -5 & 12 & 23 & 4 & -19 & -9 & 23 & 27 & -5 & -25 & -7 & 20 & 17 & -6 & -13 & -2 & 8 & 4 & -2 & -2 & 0 & 1 \\
&&&&&&&&&&&&&&&&&&&&&&&&&&&&&&&&&&&&&& \\
0 & -1 & 2 & -2 & -1 & 0 & 2 & -2 & -3 & 0 & -2 & -7 & 7 & 5 & -11 & -33 & 4 & 31 & 9 & -49 & -37 & 24 & 49 & -6 & -47 & -27 & 29 & 28 & -6 & -24 & -6 & 11 & 6 & -4 & -3 & 0 & 2 & -1 & 0 \\
&&&&&&&&&&&&&&&&&&&&&&&&&&&&&&&&&&&&&& \\
1 & -1 & 0 & 1 & 1 & -1 & 0 & 5 & -2 & -10 & 8 & 20 & 1 & -35 & -7 & 46 & 36 & -34 & -51 & 12 & 61 & 19 & -37 & -36 & 14 & 32 & 7 & -16 & -11 & 5 & 7 & 0 & -2 & -1 & 1 & 0 & 0 & 0 & 0 \\
&&&&&&&&&&&&&&&&&&&&&&&&&&&&&&&&&&&&&& \\
-1 & 1 & 0 & 0 & -2 & 3 & 0 & -7 & -1 & 14 & 6 & -22 & -20 & 23 & 32 & -11 & -44 & -8 & 34 & 25 & -18 & -27 & -1 & 17 & 8 & -6 & -7 & 1 & 2 & 1 & -1 & 0 & 0 & 0 & 0 & 0 & 0 & 0 & 0 \\
&&&&&&&&&&&&&&&&&&&&&&&&&&&&&&&&&&&&&& \\
0 & 1 & -2 & 1 & 1 & -1 & -2 & 2 & 6 & -5 & -10 & 5 & 13 & 3 & -17 & -7 & 11 & 11 & -3 & -10 & -1 & 5 & 2 & -1 & -2 & 1 & 0 & 0 & 0 & 0 & 0 & 0 & 0 & 0 & 0 & 0 & 0 & 0 & 0
\end{array}\right).
\end{equation}
\begin{equation}
\mathcal{H}_{[2,1],[3]}=\frac{1}{q^{21}A^{5}\{A\}\{Aq\}}
\left(\begin{array}{rrrrrrrrrrrrrrrrrrrrrrrrrrrrrrrr}
0 & 0 & 0 & 0 & 0 & 0 & 0 & 0 & 1 & 0 & 0 & 1 & 0 & 1 & -1 & 3 & 0 & 1 & -3 & 2 & 2 & 2 & -3 & 0 & 0 & 3 & -1 & 0 & -1 & 1 & 0 & 0 \\
&&&&&&&&&&&&&&&&&&&&&&&&&&&&&&& \\
0 & 0 & 0 & 0 & -1 & 0 & 0 & -1 & -1 & -1 & 0 & -6 & 0 & -3 & 6 & -9 & -1 & -8 & 8 & -3 & 1 & -9 & 2 & -2 & 6 & -5 & 1 & -4 & 3 & -1 & 1 & -1 \\
&&&&&&&&&&&&&&&&&&&&&&&&&&&&&&& \\
0 & 1 & -1 & 1 & 1 & -1 & 3 & -1 & 8 & -7 & 6 & -3 & 15 & -3 & 4 & -10 & 11 & 6 & 10 & -11 & -3 & 1 & 14 & 0 & -4 & -5 & 3 & 4 & 0 & -1 & -1 & 1 \\
&&&&&&&&&&&&&&&&&&&&&&&&&&&&&&& \\
-1 & 0 & 2 & -3 & 0 & -2 & 3 & -1 & -7 & 0 & -2 & 9 & -7 & -7 & -12 & 13 & 6 & -1 & -19 & -1 & 6 & 9 & -5 & -5 & -3 & 3 & 1 & 0 & -1 & 0 & 0 & 0 \\
&&&&&&&&&&&&&&&&&&&&&&&&&&&&&&& \\
1 & -1 & 0 & 1 & -1 & 4 & -4 & 3 & -6 & 9 & 2 & 0 & -13 & 5 & 9 & 7 & -10 & -6 & 3 & 6 & 2 & -3 & -1 & 0 & 1 & 0 & 0 & 0 & 0 & 0 & 0 & 0 \\
&&&&&&&&&&&&&&&&&&&&&&&&&&&&&&& \\
0 & 0 & -1 & 1 & 1 & -1 & -2 & 0 & 5 & -1 & -5 & -2 & 5 & 3 & -2 & -4 & 1 & 1 & 1 & -1 & 0 & 0 & 0 & 0 & 0 & 0 & 0 & 0 & 0 & 0 & 0 & 0
\end{array}\right).
\end{equation}
\begin{equation}
\mathcal{H}_{[2,1],[2,1]}=\frac{A^{3}}{q^{30}\{A\}\{Aq\}\{A/q\}}
\left(\begin{array}{rrrrrrrrrrrrrrrrrrrrrrrrrrrrrrr}
0 & 0 & 0 & 0 & 0 & -1 & 4 & -9 & 16 & -26 & 36 & -44 & 52 & -60 & 62 & -61 & 62 & -60 & 52 & -44 & 36 & -26 & 16 & -9 & 4 & -1 & 0 & 0 & 0 & 0 & 0 \\
&&&&&&&&&&&&&&&&&&&&&&&&&&&&&& \\
0 & 0 & 1 & -3 & 5 & -5 & 4 & 3 & -18 & 42 & -70 & 104 & -137 & 169 & -186 & 191 & -186 & 169 & -137 & 104 & -70 & 42 & -18 & 3 & 4 & -5 & 5 & -3 & 1 & 0 & 0 \\
&&&&&&&&&&&&&&&&&&&&&&&&&&&&&& \\
-1 & 4 & -11 & 21 & -38 & 60 & -86 & 107 & -128 & 138 & -143 & 127 & -115 & 100 & -89 & 75 & -89 & 100 & -115 & 127 & -143 & 138 & -128 & 107 & -86 & 60 & -38 & 21 & -11 & 4 & -1 \\
&&&&&&&&&&&&&&&&&&&&&&&&&&&&&& \\
2 & -7 & 19 & -40 & 78 & -129 & 196 & -267 & 354 & -430 & 502 & -554 & 610 & -630 & 654 & -653 & 654 & -630 & 610 & -554 & 502 & -430 & 354 & -267 & 196 & -129 & 78 & -40 & 19 & -7 & 2 \\
&&&&&&&&&&&&&&&&&&&&&&&&&&&&&& \\
-1 & 3 & -10 & 24 & -51 & 87 & -142 & 207 & -287 & 361 & -447 & 513 & -579 & 615 & -652 & 652 & -652 & 615 & -579 & 513 & -447 & 361 & -287 & 207 & -142 & 87 & -51 & 24 & -10 & 3 & -1 \\
&&&&&&&&&&&&&&&&&&&&&&&&&&&&&& \\
0 & 0 & 1 & -2 & 6 & -12 & 24 & -37 & 58 & -78 & 106 & -126 & 151 & -163 & 179 & -178 & 179 & -163 & 151 & -126 & 106 & -78 & 58 & -37 & 24 & -12 & 6 & -2 & 1 & 0 & 0 \\
&&&&&&&&&&&&&&&&&&&&&&&&&&&&&& \\
0 & 0 & 0 & 0 & 0 & -1 & 2 & -5 & 9 & -15 & 20 & -27 & 32 & -38 & 40 & -42 & 40 & -38 & 32 & -27 & 20 & -15 & 9 & -5 & 2 & -1 & 0 & 0 & 0 & 0 & 0
\end{array}\right).
\end{equation}
}}
\end{landscape}
\newpage

\setlength{\arraycolsep}{6pt}




\begin{thebibliography}{99}



\bibitem{Wit} E. Witten,
Comm.Math.Phys. {\bf 121} (1989)  351-399

\bibitem{Jones} V.F.R. Jones, 
Invent.Math. {\bf 72} (1983) 1
Bull.AMS {\bf 12} (1985) 103
Ann.Math. {\bf 126} (1987) 335\\
L. Kauffman, 
Topology, {\bf 26} (1987) 395

\bibitem{CS} S.-S. Chern, J. Simons,
Ann.Math. {\bf 99} (1974) 48-69

\bibitem{HOMFLY} P. Freyd, D. Yetter, J. Hoste, W.B.R. Lickorish, K. Millet,
A. Ocneanu,
Bull. AMS. {\bf 12} (1985) 239\\
J.H. Przytycki, K.P. Traczyk, 
Kobe J. Math. {\bf 4} (1987) 115-139

\bibitem{Kauf} L. Kauffman, Transactions of the American Mathematical Society, {\bf 318} (1990) 417–471

\bibitem{inds} R.K. Kaul, T.R. Govindarajan, Nucl.Phys. {\bf B380} (1992)
293-336, hep-th/9111063;\\
P. Ramadevi, T.R. Govindarajan, R.K. Kaul, Nucl.Phys. {\bf B402} (1993)
548-566, hep-th/9212110;
Nucl.Phys. {\bf B422} (1994) 291-306, hep-th/9312215;\\
P. Ramadevi, T. Sarkar,
Nucl.Phys. {\bf B600} (2001) 487-511,
hep-th/0009188\\
Zodinmawia and P. Ramadevi, Nucl.Phys. {\bf B870} (2013) 205-242, arXiv:1107.3918;  arXiv:1209.1346

\bibitem{OV} H. Ooguri, C. Vafa, Nucl.Phys. {\bf B577} (2000) 419-438, arXiv:hep-th/9912123

\bibitem{LMOV} J.M.F. Labastida, M. Mari\~no, Commun.Math.Phys. {\bf 217} (2001) 423-449, hep-th/0004196; math/0104180\\
J.M.F. Labastida, M. Mari\~no, C. Vafa, JHEP {\bf 0011} (2000) 007, hep-th/0010102\\
M. Mari\~no and C. Vafa, hep-th/0108064\\
K. Liu and P. Peng,
J. Diff. Geom. {\bf 85} (2010), no. 3 479-525, arXiv:0704.1526;
Math.Res.Lett. {\bf 17} (2010) 493-506, arXiv:1012.2635\\
M. Marino,
Commun.Math.Phys. {\bf 298} (2010) 613�643, arXiv:0904.1088\\
S. Stevan, Annales Henri Poincare {\bf 11} (2010) 1201-1224, arXiv:1003.2861\\
C. Paul, P. Borhade and P. Ramadevi, arXiv:1003.5282; Nucl.Phys. {\bf B841} (2010) 448-462, arXiv:1008.3453\\
S. Nawata, P. Ramadevi and Zodinmawia, JHEP {\bf 1401} (2014) 126, arXiv:1310.2240\\
S. Garoufalidis, P. Kucharski, P. Su\l kowski, Commun.Math.Phys. {\bf 346} (2016) 75-113, arXiv:1504.06327\\
P. Kucharski, P. Su\l kowski, JHEP {\bf 11} (2016) 120, arXiv:1608.06600\\
Wei Luo, Shengmao Zhu, arXiv:1611.06506\\
A. Mironov, A. Morozov, An. Morozov, P. Ramadevi, Vivek Kumar Singh, A. Sleptsov, arXiv:1702.06316\\
M. Kameyama, S. Nawata, arXiv:1703.05408\\
A. Mironov, A. Morozov, An. Morozov, A. Sleptsov, Nucl.Phys. {\bf B924} (2017) 1-32, arXiv:1706.00761\\
P. Kucharski, M. Reineke, M. Stosic, P. Su\l kowski, Phys.Rev. {\bf D96} (2017) 121902, arXiv:1707.02991; arXiv:1707.04017

\bibitem{QC} D. Melnikov, A. Mironov, S. Mironov, A. Morozov, An. Morozov, arXiv:1703.00431

\bibitem{MMfing} A. Mironov, A. Morozov, Nucl.Phys. {\bf B899} (2015) 395-413,
arXiv:1506.00339

\bibitem{Rama2} A. Mironov, A. Morozov, An. Morozov, P. Ramadevi, Vivek Kumar Singh, A. Sleptsov, J.Phys. {\bf A50} (2017) 085201, arXiv:1601.04199

\bibitem{MMMSint} A. Mironov, A. Morozov, An. Morozov, A. Sleptsov, Phys.Lett. {\bf B760} (2016) 45-58, arXiv:1605.04881

\bibitem{RacahRama} S. Nawata, P. Ramadevi and Zodinmawia, Lett.Math.Phys. {\bf 103} (2013) 1389-1398, arXiv:1302.5143

\bibitem{RT} E. Guadagnini, M. Martellini, M. Mintchev, Clausthal 1989,
Proceedings, {\sl Quantum groups},
307-317;
Phys.Lett. {\bf B235} (1990) 275\\
N.Yu. Reshetikhin, V.G. Turaev, 
Comm.Math.Phys. {\bf 127} (1990) 1-26

\bibitem{MMMI} A. Mironov, A. Morozov, An. Morozov,
JHEP 1203 (2012) 034, arXiv:1112.2654

\bibitem{RTmod} A. Mironov, A. Morozov, An. Morozov,
in: {\it Strings, Gauge Fields, and the Geometry Behind: The Legacy of Maximilian Kreuzer},
eds: A.Rebhan, L.Katzarkov, J.Knapp, R.Rashkov, E.Scheidegger,
World Scietific, 2013 pp.101-118,
arXiv:1112.5754\\
H. Itoyama, A. Mironov, A. Morozov, An. Morozov,
Int.J.Mod.Phys. {\bf A27} (2012) 1250099,
arXiv:1204.4785 \\
A. Anokhina, A. Mironov, A. Morozov, An. Morozov,
Nucl.Phys. {\bf B868} (2013) 271-313, arXiv:1207.0279;
Adv.High Energy Phys. 2013 (2013) 931830,
arXiv:1304.1486  \\
A. Anokhina, arXiv:1412.8444\\
Saswati Dhara, A. Mironov, A. Morozov, An. Morozov, P. Ramadevi, Vivek Kumar Singh, A. Sleptsov, arXiv:1711.10952

\bibitem{MMMS21}
A. Mironov, A. Morozov, An. Morozov, A. Sleptsov, J. Mod. Phys. {\bf A30} (2015) 1550169, arXiv:1508.02870

\bibitem{MMMS31} A. Mironov, A. Morozov, An. Morozov, A. Sleptsov, JHEP, {\bf 2016} (2016) 134,  arXiv:1605.02313; JETP Lett. {\bf 104} (2016) 56-61,
arXiv:1605.03098\\
Sh. Shakirov, A. Sleptsov, arXiv:1611.03797

\bibitem{IMMMec} H. Itoyama, A. Mironov, A. Morozov, An. Morozov,
Int.J.Mod.Phys. {\bf A28} (2013) 1340009,
arXiv:1209.6304 \\
A. Mironov, A. Morozov, arXiv:1610.03043

\bibitem{Univ} A. Mironov, A. Morozov, Phys.Lett. {\bf B755} (2016) 47-57, arXiv:1511.09077

\bibitem{BJLMMMS} C. Bai, J. Jiang, J. Liang, A. Mironov, A. Morozov, An. Morozov, A. Sleptsov, arXiv:1709.09228

\bibitem{Cau} J.H. Conway, 
Algebraic Properties,
In: John Leech (ed.), {\sl Computational Problems in Abstract Algebra}, Proc.
Conf.
Oxford, 1967, Pergamon Press, Oxford-New York, 329-358, 1970\\
A. Caudron, {\it Classification des noeuds et des enlacements}, Publ. Math. Orsay
{\bf 82-4}, University of Paris XI, Orsay, 1982\\
F. Bonahon, L.C. Siebenmann,  http://www-bcf.usc.edu/$\sim$fbonahon/Research/Preprints/BonSieb.pdf, {\it New geometric splittings of classical knots and the classification and symmetries of arborescent knots}, 2010

\bibitem{arbor}
P. Ramadevi, T.R. Govindarajan, R.K. Kaul, Mod.Phys.Lett. {\bf A9} (1994) 3205-3218, hep-th/9401095\\
S. Nawata, P. Ramadevi, Zodinmawia,
J.Knot Theory and Its Ramifications {\bf 22} (2013) 13, arXiv:1302.5144\\
D. Galakhov, D. Melnikov, A. Mironov, A. Morozov, A. Sleptsov, Phys.Lett. {\bf B743} (2015) 71-74, arXiv:1412.2616\\
Zodinmawia's PhD thesis, 2014\\
A. Mironov, A. Morozov, A. Sleptsov, JHEP {\bf 07} (2015) 069,  arXiv:1412.8432\\
S. Nawata, P. Ramadevi, Vivek Kumar Singh,  J.Knot Theor.Ramifications {\bf 26} (2017) 1750096, arXiv:1504.00364

\bibitem{MMSpret} A. Mironov, A. Morozov, A. Sleptsov, JHEP {\bf 07} (2015) 069,  arXiv:1412.8432

\bibitem{evo} P. Dunin-Barkowski, A. Mironov, A. Morozov, A. Sleptsov,
A. Smirnov, JHEP {\bf 03} (2013) 021,
arXiv:1106.4305\\
A. Mironov, A. Morozov, An. Morozov,
AIP Conf.Proc. {\bf 1562} (2013) 123-155, arXiv:1306.3197\\
S. Arthamonov, A. Mironov, A. Morozov, An. Morozov,  JHEP {\bf 04} (2014) 156,  arXiv:1309.7984

\bibitem{Rama1} A. Mironov, A. Morozov, An. Morozov, P. Ramadevi,
V.K.Singh,
JHEP {\bf 1507} (2015) 109,  arXiv:1504.00371

\bibitem{AMcabling} A. Anokhina, An. Morozov, Teor.Mat.Fiz. {\bf 178} (2014) 3-68,
arXiv:1307.2216

\bibitem{GJ} J. Gu, H. Jockers, arXiv:1407.5643


\bibitem{knotebook} http://knotebook.org

\bibitem{Wentzl} I. Tuba, H. Wenzl, math/9912013

\bibitem{twi} http://katlas.org/wiki/The\_Thistlethwaite\_Link\_Table

\bibitem{Inds8} S. Nawata, P. Ramadevi, Zodinmawia, X. Sun, JHEP {\bf 1211} (2012) 157, arXiv:1209.1409



\end{thebibliography}
\end{document}